\documentclass[11pt]{article}

\usepackage{amssymb}
\usepackage{amsmath}
\usepackage{latexsym}

\newtheorem{theorem}{Theorem}[section]
\newtheorem{lemma}[theorem]{Lemma}
\newtheorem{conjecture}{Conjecture}

\def\qed{\hfill $\Box$\vspace{2ex}}
\def\vertex{\circle*{1.5}}
\begin{document}

\title{{\bf A characterization of b-perfect graphs}\thanks{Supported
by Centre Jacques Cartier.}}

\author{%
{\bf Ch\'{\i}nh T. Ho\`ang}\thanks{%
Dept.~of Physics and Computer Science, Wilfrid Laurier University,
75 University Avenue West, Waterloo, Ontario, Canada N2L 3C5.
Supported by NSERC.}
\and%
{\bf Fr\'ed\'eric Maffray}\thanks{%
C.N.R.S, Laboratoire G-SCOP, 46 Avenue F\'elix Viallet, 38031 Grenoble
Cedex, France.}
\and%
{\bf Meriem Mechebbek}\thanks{USTHB, Laboratoire LAID3, BP32 El Alia,
Bab Ezzouar 16111, Alger, Algeria.}}

\date{\today}

\maketitle

\begin{abstract}
A b-coloring is a coloring of the vertices of a graph such that each
color class contains a vertex that has a neighbor in all other color
classes, and the b-chromatic number of a graph $G$ is the largest
integer $k$ such that $G$ admits a b-coloring with $k$ colors.  A
graph is b-perfect if the b-chromatic number is equal to the chromatic
number for every induced subgraph of $G$.  We prove that a graph is
b-perfect if and only if it does not contain as an induced subgraph a
member of a certain list of twenty-two graphs.  This entails the
existence of a polynomial-time recognition algorithm and of a
polynomial-time algorithm for coloring exactly the vertices of every
b-perfect graph.


\end{abstract}

\section{Introduction}

A proper coloring of a graph $G$ is a mapping $c$ from the vertex-set
$V(G)$ of $G$ to the set $\{1, 2, \ldots\}$ of positive integers
(colors) such that any two adjacent vertices are mapped to different
colors.  Each set of vertices colored with one color is a stable set
of vertices of $G$, so a coloring is a partition of $V(G)$ into stable
sets.  The smallest number $k$ for which $G$ admits a coloring with
$k$ colors is the chromatic number $\chi(G)$ of $G$ \cite{berg}.

Many graph invariants related to colorings have been defined.  Most of
them try to minimize the number of colors used to color the vertices
under some constraints.  For some other invariants, it is meaningful
to try to maximize this number.  The b-chromatic number is such an
example.  When we try to color the vertices of a graph, we can start
from a given coloring and try to decrease the number of colors by
eliminating color classes.  One possible such procedure consists in
trying to reduce the number of colors by transferring every vertex
from a fixed color class to a color class in which it has no
neighbour, if any such class exists.  A \emph{b-coloring} is a proper
coloring in which this is not possible, that is, every color class $i$
contains at least one vertex that has a neighbor in all the other
classes.  Any such vertex will be called a \emph{b-vertex} of color
$i$.  The \emph{b-chromatic number} $b(G)$ is the largest integer $k$
such that $G$ admits a b-coloring with $k$ colors.  Irving and Manlove
\cite{im, m} proved that deciding whether a graph $G$ admits a
b-coloring with a given number of colors is an NP-complete problem,
even when it is restricted to the class of bipartite graphs
\cite{KTV}.  On the other hand, they gave a polynomial-time algorithm
that solves this problem for trees.  The NP-completeness results has
incited researchers to establish bounds on the b-chromatic number in
general or to find its exact values for subclasses of graphs.

\begin{figure}[ht]
\unitlength=0.06cm
\begin{center}
\begin{tabular}{|c|c|c|c|c|}
       \hline
%\begin{tabular}{ccc}
\begin{picture}(16,26) % GRAPH F1
      % vertices
\multiput(0,12)(8,0){3}{\vertex}
\multiput(4,20)(8,0){2}{\vertex}
      % edges
\multiput(0,12)(8,0){2}{\line(1,2){4}}
\multiput(8,12)(8,0){2}{\line(-1,2){4}}
\put(5,2){$F_1$}
      % end
\end{picture}
&
%\hspace{1cm}
\begin{picture}(20,26) % GRAPH F2
      % vertices
\multiput(0,12)(8,0){2}{\vertex}
\multiput(12,12)(8,0){2}{\vertex}
\put(4,20){\vertex}
\multiput(12,20)(8,0){2}{\vertex}
      % edges
\put(0,12){\line(1,2){4}} \put(8,12){\line(-1,2){4}}
\multiput(12,12)(8,0){2}{\line(0,1){8}}
\put(12,20){\line(1,0){8}}
\put(7,2){$F_2$}
      % end
\end{picture}
%\hspace{1cm}
&
\begin{picture}(32,26) % GRAPH F3
      % vertices
\multiput(0,12)(12,0){3}{\vertex}
\multiput(8,12)(12,0){3}{\vertex}
\multiput(4,20)(12,0){3}{\vertex}
      % edges
\multiput(0,12)(12,0){3}{\line(1,2){4}}
\multiput(8,12)(12,0){3}{\line(-1,2){4}}
\put(13,2){$F_3$}
      % end
\end{picture}
&
\begin{picture}(24,26) % GRAPH F4
      % vertices
\multiput(8,12)(8,0){2}{\vertex}
\multiput(8,20)(8,0){2}{\vertex}
\multiput(0,16)(24,0){2}{\vertex}
      % edges
\multiput(8,12)(0,8){2}{\line(1,0){8}}
\multiput(8,12)(8,0){2}{\line(0,1){8}}
\multiput(0,16)(16,4){2}{\line(2,-1){8}}
\multiput(0,16)(16,-4){2}{\line(2,1){8}}
\put(8,12){\line(1,1){8}}
\put(9,2){$F_4$}
      % end
\end{picture}
&
% $F_1=P_5$ & $F_2=P_3+P_4$ & $F_3=3P_3$ & $F_4$ \\
%
\begin{picture}(24,30) % GRAPH F5
      % vertices
\multiput(6,12)(12,0){2}{\vertex}
\multiput(0,18)(12,0){3}{\vertex}
\multiput(6,24)(12,0){2}{\vertex}
      % edges
\put(12,18){\line(1,0){12}}
\put(6,12){\line(0,1){12}}
\multiput(0,18)(6,6){2}{\line(1,-1){6}}
\multiput(12,18)(6,6){2}{\line(1,-1){6}}
\multiput(0,18)(6,-6){2}{\line(1,1){6}}
\multiput(12,18)(6,-6){2}{\line(1,1){6}}
\put(10,2){$F_5$}
      % end
\end{picture}
\\ \hline
\begin{picture}(30,30) % GRAPH F6
      % vertices
\multiput(6,12)(18,0){2}{\vertex}
\multiput(0,18)(18,0){2}{\vertex}
\multiput(12,18)(18,0){2}{\vertex}
\multiput(6,24)(18,0){2}{\vertex}
      % edges
\multiput(0,18)(18,0){2}{\line(1,0){12}}
\multiput(0,18)(18,0){2}{\line(1,-1){6}}
\multiput(6,24)(18,0){2}{\line(1,-1){6}}
\multiput(6,12)(18,0){2}{\line(1,1){6}}
\multiput(0,18)(18,0){2}{\line(1,1){6}}
\put(12,2){$F_6$}
      % end
\end{picture}
&
\begin{picture}(30,30) % GRAPH F7
      % vertices
\multiput(6,12)(18,0){2}{\vertex}
\multiput(0,18)(18,0){2}{\vertex}
\multiput(12,18)(18,0){2}{\vertex}
\multiput(6,24)(18,0){2}{\vertex}
      % edges
\put(0,18){\line(1,0){30}}
\multiput(0,18)(18,0){2}{\line(1,-1){6}}
\multiput(6,24)(18,0){2}{\line(1,-1){6}}
\multiput(6,12)(18,0){2}{\line(1,1){6}}
\multiput(0,18)(18,0){2}{\line(1,1){6}}
\put(12,2){$F_7$}
      % end
\end{picture}
&
\begin{picture}(18,39) % GRAPH F8
      % vertices
\multiput(0,12)(6,0){4}{\vertex}
\multiput(9,21)(0,6){2}{\vertex}
\multiput(3,33)(12,0){2}{\vertex}
      % edges
\put(0,12){\line(1,0){18}}
\put(3,33){\line(1,0){12}}
\put(9,21){\line(0,1){6}}
\put(0,12){\line(1,1){9}}\put(18,12){\line(-1,1){9}}
\put(6,12){\line(1,3){3}}\put(12,12){\line(-1,3){3}}
\put(9,27){\line(1,1){6}}
\put(9,27){\line(-1,1){6}}
\put(6,2){$F_8$}
      % end
\end{picture}
&
\begin{picture}(30,39) % GRAPH F9
      % vertices
\multiput(0,12)(6,0){6}{\vertex}
\multiput(15,21)(0,6){2}{\vertex}
\multiput(9,33)(12,0){2}{\vertex}
      % edges
\multiput(0,12)(18,0){2}{\line(1,0){12}}
\put(9,33){\line(1,0){12}}
\put(15,21){\line(0,1){6}}
\put(0,12){\line(5,3){15}}\put(30,12){\line(-5,3){15}}
\put(6,12){\line(1,1){9}}\put(24,12){\line(-1,1){9}}
\put(12,12){\line(1,3){3}}\put(18,12){\line(-1,3){3}}
\put(15,27){\line(1,1){6}}
\put(15,27){\line(-1,1){6}}
\put(12,2){$F_9$}
      % end
\end{picture}
&
\begin{picture}(27,33) % GRAPH F10
      % vertices
\multiput(0,12)(27,0){2}{\vertex}
\multiput(9,18)(9,0){2}{\vertex}
\multiput(9,27)(9,0){2}{\vertex}
      % edges
\put(0,12){\line(1,0){27}}
\multiput(9,18)(0,9){2}{\line(1,0){9}}
\multiput(9,18)(9,0){2}{\line(0,1){9}}
\put(0,12){\line(3,2){9}} \put(27,12){\line(-3,2){9}}
\put(0,12){\line(3,5){9}} \put(27,12){\line(-3,5){9}}
\put(9,18){\line(1,1){9}}
\put(10,2){$F_{10}$}
      % end
\end{picture}
\\
\hline
\begin{picture}(24,30) % GRAPH F11
      % vertices
\multiput(6,12)(12,0){2}{\vertex}
\multiput(0,18)(12,0){3}{\vertex}
\multiput(6,24)(12,0){2}{\vertex}
      % edges
\put(12,18){\line(1,0){12}}
\put(6,12){\line(0,1){12}}
\multiput(0,18)(6,6){2}{\line(1,-1){6}}
\multiput(12,18)(6,6){2}{\line(1,-1){6}}
\multiput(0,18)(6,-6){2}{\line(1,1){6}}
\multiput(12,18)(6,-6){2}{\line(1,1){6}}
\put(0,18){\line(3,1){18}} \put(0,18){\line(3,-1){18}}
\put(10,2){$F_{11}$}
      % end
\end{picture}
&
\begin{picture}(27,27) % GRAPH F12
      % vertices
\multiput(0,12)(9,0){4}{\vertex}
\multiput(0,21)(9,0){4}{\vertex}
      % edges
\multiput(0,12)(0,9){2}{\line(1,0){27}}
\multiput(9,12)(9,0){2}{\line(0,1){9}}
\multiput(0,12)(18,0){2}{\line(1,1){9}}
\multiput(9,12)(18,0){2}{\line(-1,1){9}}
\put(10,2){$F_{12}$}
      % end
\end{picture}
&
\begin{picture}(27,27) % GRAPH F13
      % vertices
\multiput(0,12)(9,0){4}{\vertex}
\multiput(0,21)(9,0){4}{\vertex}
      % edges
\multiput(0,12)(0,9){2}{\line(1,0){27}}
\multiput(0,12)(27,0){2}{\line(0,1){9}}
\multiput(0,12)(9,0){3}{\line(1,1){9}}
\multiput(9,12)(9,0){3}{\line(-1,1){9}}
\put(10,2){$F_{13}$}
      % end
\end{picture}
&
\begin{picture}(27,30) % GRAPH F14
      % vertices
\multiput(0,12)(9,0){4}{\vertex}
\multiput(9,21)(9,0){2}{\vertex}
\multiput(0,24)(27,0){2}{\vertex}
      % edges
\multiput(0,12)(0,12){2}{\line(1,0){27}}
\put(9,21){\line(1,0){9}}
\multiput(0,12)(27,0){2}{\line(0,1){12}}
\multiput(0,12)(9,0){2}{\line(1,1){9}}
\multiput(18,12)(9,0){2}{\line(-1,1){9}}
\put(9,21){\line(-3,1){9}} \put(18,21){\line(3,1){9}}
\put(9,12){\line(-3,4){9}} \put(18,12){\line(3,4){9}}
\put(10,2){$F_{14}$}
      % end
\end{picture}
&
\begin{picture}(27,33) % GRAPH F15
      % vertices
\multiput(0,12)(27,0){2}{\vertex}
\multiput(9,15)(9,0){2}{\vertex}
\multiput(9,24)(9,0){2}{\vertex}
\multiput(0,27)(27,0){2}{\vertex}
      % edges
\multiput(0,12)(0,15){2}{\line(1,0){27}}
\multiput(9,15)(0,9){2}{\line(1,0){9}}
\multiput(0,12)(27,0){2}{\line(0,1){15}}
\put(9,15){\line(1,1){9}} \put(18,15){\line(-1,1){9}}
\multiput(0,12)(18,3){2}{\line(3,4){9}}
\multiput(9,15)(18,-3){2}{\line(-3,4){9}}
\multiput(0,12)(18,12){2}{\line(3,1){9}}
\multiput(9,24)(18,-12){2}{\line(-3,1){9}}
\put(10,2){$F_{15}$}
      % end
\end{picture}
\\
\hline
\begin{picture}(24,36) % GRAPH F16
      % vertices
\multiput(6,12)(12,0){2}{\vertex}
\multiput(0,24)(24,0){2}{\vertex}
\multiput(12,21)(0,9){2}{\vertex}
      % edges
\put(6,12){\line(1,0){12}}
\put(12,21){\line(0,1){9}}
\put(18,12){\line(1,2){6}} \put(6,12){\line(-1,2){6}}
\put(12,21){\line(4,1){12}} \put(12,21){\line(-4,1){12}}
\put(12,30){\line(2,-1){12}} \put(12,30){\line(-2,-1){12}}
\put(9,2){$F_{16}$}
      % end
\end{picture}
&
\begin{picture}(24,36) % GRAPH F17
      % vertices
\multiput(6,12)(12,0){2}{\vertex}
\multiput(0,24)(24,0){2}{\vertex}
\multiput(12,21)(0,9){2}{\vertex}
      % edges
\put(6,12){\line(1,0){12}}
\put(12,21){\line(0,1){9}}
\put(18,12){\line(1,2){6}} \put(6,12){\line(-1,2){6}}
\put(12,21){\line(4,1){12}} \put(12,21){\line(-4,1){12}}
\put(12,30){\line(2,-1){12}} \put(12,30){\line(-2,-1){12}}
\put(12,21){\line(2,-3){6}}
\put(9,2){$F_{17}$}
      % end
\end{picture}
&
\begin{picture}(24,36) % GRAPH F18
      % vertices
\multiput(6,12)(12,0){2}{\vertex}
\multiput(6,18)(12,0){2}{\vertex}
\multiput(0,24)(12,0){3}{\vertex}
\put(12,30){\vertex}
      % edges of the C5
\put(6,12){\line(1,0){12}}
\put(18,12){\line(1,2){6}} \put(6,12){\line(-1,2){6}}
\put(12,30){\line(2,-1){12}} \put(12,30){\line(-2,-1){12}}
      % Other edges
\multiput(6,12)(12,0){2}{\line(0,1){6}}
\put(6,18){\line(3,1){18}} \put(18,18){\line(-3,1){18}}
\put(6,18){\line(1,1){6}} \put(18,18){\line(-1,1){6}}
\put(6,18){\line(-1,1){6}} \put(18,18){\line(1,1){6}}
\put(9,2){$F_{18}$}
      % end
\end{picture}
&
\begin{picture}(24,36) % GRAPH F19
      % vertices
\multiput(6,12)(12,0){2}{\vertex}
\multiput(3,21)(18,0){2}{\vertex}
\multiput(0,24)(24,0){2}{\vertex}
\multiput(12,27)(0,3){2}{\vertex}
      % edges
\put(6,12){\line(1,0){12}}\put(3,21){\line(1,0){18}}
\put(6,12){\line(-1,3){3}} \put(18,12){\line(1,3){3}}
\put(6,12){\line(-1,2){6}} \put(18,12){\line(1,2){6}}
\put(3,21){\line(3,2){9}} \put(21,21){\line(-3,2){9}}
\put(3,21){\line(1,1){9}} \put(21,21){\line(-1,1){9}}
\put(0,24){\line(4,1){12}} \put(24,24){\line(-4,1){12}}
\put(0,24){\line(2,1){12}} \put(24,24){\line(-2,1){12}}
\put(9,2){$F_{19}$}
      % end
\end{picture}
&
\begin{picture}(24,36) % GRAPH F20
      % vertices
\multiput(6,12)(12,0){2}{\vertex}
\multiput(6,18)(12,0){2}{\vertex}
\multiput(0,24)(12,0){3}{\vertex}
\put(12,30){\vertex}
      % edges
\multiput(6,12)(0,6){2}{\line(1,0){12}}
\put(18,12){\line(1,2){6}} \put(6,12){\line(-1,2){6}}
\multiput(6,18)(6,12){2}{\line(2,-1){12}}
\multiput(6,12)(-6,12){2}{\line(2,1){12}}
\put(0,24){\line(1,0){24}}
\put(6,18){\line(-1,1){6}} \put(18,18){\line(1,1){6}}
\put(6,18){\line(1,1){6}}
\put(18,18){\line(-1,2){6}}
\put(9,2){$F_{20}$}
      % end
\end{picture}
\\ \hline
%%%
\begin{picture}(24,36) % GRAPH F21
      % vertices
\multiput(6,12)(12,0){2}{\vertex}
\multiput(6,18)(12,0){2}{\vertex}
\multiput(0,24)(12,0){3}{\vertex}
\put(12,30){\vertex}
      % edges
\multiput(6,12)(0,6){2}{\line(1,0){12}}
\put(18,12){\line(1,2){6}} \put(6,12){\line(-1,2){6}}
\multiput(6,18)(6,12){2}{\line(2,-1){12}}
\multiput(6,12)(-6,12){2}{\line(2,1){12}}
\put(0,24){\line(1,0){24}}
\put(6,18){\line(-1,1){6}} \put(18,18){\line(1,1){6}}
\put(6,18){\line(1,1){6}}
\put(18,18){\line(-1,2){6}}
\put(6,12){\line(1,2){6}}
\put(9,2){$F_{21}$}
      % end
\end{picture}
&
\begin{picture}(24,36) % GRAPH F22
      % vertices
\multiput(6,12)(12,0){2}{\vertex}
\multiput(0,24)(8,0){4}{\vertex}
\multiput(12,15)(0,15){2}{\vertex}
      % edges
\put(6,12){\line(1,0){12}}
\put(0,24){\line(1,0){24}}
\put(12,15){\line(0,1){15}}
\put(18,12){\line(1,2){6}} \put(6,12){\line(-1,2){6}}
\put(12,30){\line(2,-1){12}} \put(12,30){\line(-2,-1){12}}
\put(6,12){\line(5,6){10}} \put(18,12){\line(-5,6){10}}
\put(6,12){\line(2,1){6}} \put(18,12){\line(-2,1){6}}
\put(8,24){\line(2,3){4}} \put(16,24){\line(-2,3){4}}
\put(9,2){$F_{22}$}
      % end
\end{picture}
&
% EMPTY BOX
&
% EMPTY BOX
&
% EMPTY BOX
\\
\hline
\end{tabular}
\end{center}
\caption{Class ${\cal F}=\{F_1, \ldots, F_{22}\}$}
\label{fig1}
\end{figure}
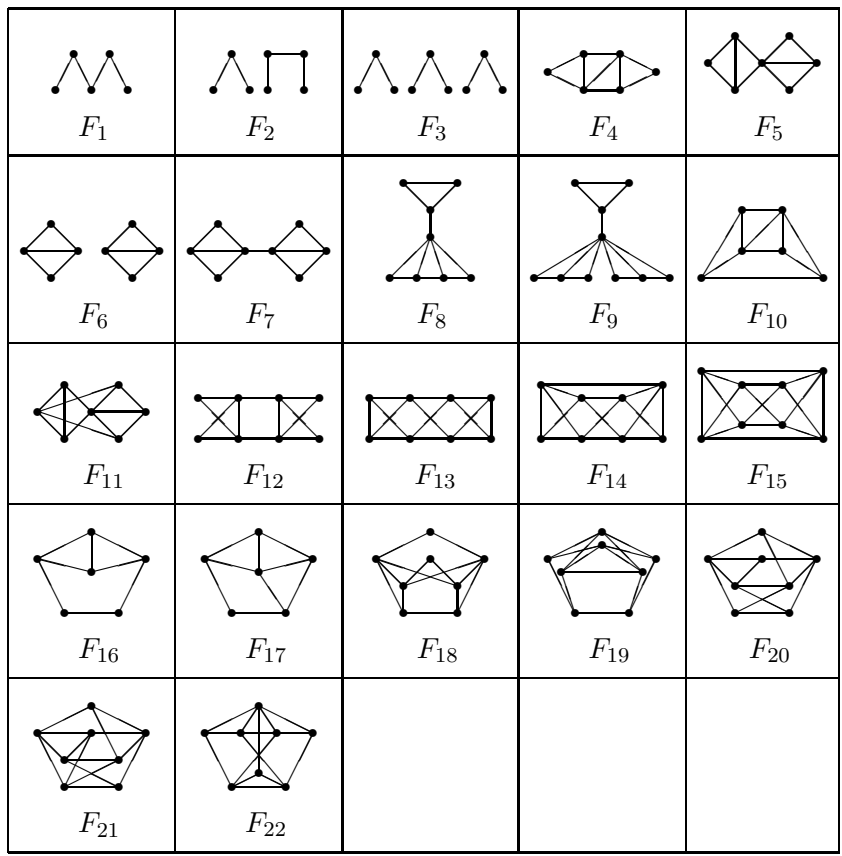

Clearly every $\chi(G)$-coloring of a graph $G$ is a b-coloring, and
so every graph $G$ satisfies $\chi(G)\leq b(G)$.  As usual with such
an inequality, it may be interesting to look at the graphs that
satisfy it with equality.  However, graphs such that $\chi(G)= b(G)$
do not have a specific structure; to see this, we can take any
arbitrary graph $G$ and add a component that consists of a clique of
size $b(G)$; we obtain a graph $G'$ that satisfies $\chi(G')= b(G')
=b(G)$.  This led Ho\`ang and Kouider \cite{hoakou} to introduce the
class of \emph{b-perfect} graphs: a graph $G$ is called b-perfect if
every induced subgraph $H$ of $G$ satisfies $\chi(H)=b(H)$.  Ho\`ang
and Kouider \cite{hoakou} proved the b-perfectness of some classes of
graphs, and asked for a good characterization of the whole class of
b-perfect graphs.  Ho\`ang, Linhares Sales and Maffray
\cite{HLM} proposed the conjecture below.  Here we solve the
problem by establishing the validity of the conjecture.  For a fixed
graph $F$, we say that a graph $G$ is \emph{$F$-free} if it does not
have an induced subgraph that is isomporphic to $F$.  For a set ${\cal
F}$ of graphs, we say that a graph $G$ is \emph{${\cal F}$-free} if it
does not have an induced subgraph that is isomporphic to a member of
${\cal F}$.  Let ${\cal{F}} = \{F_1, \ldots, F_{22}\}$ be the set of
graphs depicted in Figure~\ref{fig1}.

\begin{conjecture}[Ho\`ang, Linhares Sales, Maffray
\cite{HLM}]\label{cj1} A graph is b-perfect if and only if it is
${\cal{F}}$-free.
\end{conjecture}

Our main result is the following.

\begin{theorem}\label{thm:main}
Conjecture~\ref{cj1} is true.
\end{theorem}

The following theorem was proved before Conjecture~\ref{cj1} was
formulated, but it can be seen as evidence for its validity.
\begin{theorem}[Ho\`ang and Kouider \cite{hoakou}]\label{thm:bip}
    { \ } \\
A bipartite graph is b-perfect if and only if it contains no $F_1,
F_2$ or $F_3$. \\
A $P_4$-free graph is b-perfect if and only if it contains no $F_3$ or
$F_6$.
\end{theorem}
Moreover, some other partial results were obtained.
\begin{theorem}[Ho\`ang, Linhares Sales, Maffray \cite{HLM}]
    Conjecture~\ref{cj1} holds for $3$-colorable graphs and for
    diamond-free graphs.
\end{theorem}
\begin{theorem}[Maffray, Mechebbek \cite{mafmech}]
    Conjecture~\ref{cj1} holds for chordal graphs.
\end{theorem}

%%% DEFINITIONS

In the remainder of this section, we introduce some definitions and
notation.  For any vertex $v$ of a graph $G$, the \emph{neighborhood}
of $v$ is the set $N(v)=\{u\in V(G)\mid uv\in E\}$ and the
\emph{degree} of $v$ is $\mbox{deg}(v) = |N(v)|$.  If a vertex $x$ is
adjacent to a vertex $y$, then we will say that $x$ {\it sees} $y$,
otherwise we will say $x$ {\it misses} $y$.  Let us say that a set $A$
of vertices is \emph{complete} (respectively, \emph{anti-complete}) to
a set $B$ if every vertex of $A$ sees (respectively, misses) every
vertex of $B$.

A \emph{component} of a graph $G$ is a maximal connected subgraph of
$G$, and a \emph{co-component} of $G$ is a component of its
complementary graph.  Two vertices $x,y$ are {\it twins} if $N(x)
-\{x,y\} = N(y) -\{x,y\}$; in addition, if $x$ sees $y$ then they are
{\it true twins}, otherwise they are {\it false twins}.  A vertex $x$
{\it dominates} a vertex $y$ if $N(y) \subseteq N(x) \cup \{x\}$; $x$
and $y$ are {\it comparable} if $x$ dominates $y$, or vice versa.

For any integer $k\ge 1$, we denote by $P_k$ the chordless path with
$k$ vertices.  For integer $k\ge 3$, we denote by $C_k$ the chordless
cycle with $k$ vertices.  A \emph{diamond} is a graph with four
vertices that consists in a clique minus an edge.

%%%%
%%%% SECTION: SOME LEMMAS
%%%%

\section{Some lemmas}

We say that a graph $G$ is \emph{b-imperfect} if it is not b-perfect,
and \emph{minimally b-imperfect} if $G$ is b-imperfect and every
proper induced subgraph of $G$ is b-perfect.  We say that a graph $G$
is a \emph{minimal counterexample} to Conjecture~\ref{cj1} if it is a
counterexample (i.e., a b-imperfect ${\cal F}$-free graph) with the
smallest number of vertices and, among all such graphs, with the
smallest number of edges.  Note that every minimal counterexample is
minimally b-imperfect.

Let $\omega(G)$ denote
the number of vertices in a largest clique of $G$.

%%% LEMMA ABOUT NO CLIQUE COMPONENT
%%%
\begin{lemma}[Ho\`ang and Kouider \cite{hoakou}]
\label{lem:nok}
Let $G$ be a minimally b-imperfect graph.  Then no component of $G$ is a
clique.
\end{lemma}

%%% LEMMA ABOUT SIMPLICIAL VERTICES
%%%
\begin{lemma}[\cite{HLM}]
\label{lem:simp}
Let $G$ be graph and $x$ be any simplicial vertex of $G$.  Let $c$ be
any b-coloring of $G$ with $k$ colors, where $k > \omega(G)$.  Then
$x$ is not a b-vertex for $c$.
\end{lemma}
\emph{Proof.} Suppose that $x$ is a b-vertex for $c$.  Then all $k$
colors of $c$ appear in the clique formed by $x$ and its neighbours.
Thus $\omega(G)\ge k$, a contradiction.  $\Box$

%%% LEMMA ABOUT COMPARABLE (NON-ADJACENT) VERTICES
%%%
\begin{lemma}[\cite{HLM}]
\label{lem:compuv}
Let $G$ be a minimally b-imperfect graph, and let $c$ be any
b-coloring with $b(G)$ colors.  Let $u,v$ be two non-adjacent vertices
of $G$ such that $N(u)\subseteq N(v)$.  Then $c(u)\neq c(v)$, and $u$
is not a b-vertex.  In particular, if $N(u)=N(v)$, then none of $u,v$
is a b-vertex.
\end{lemma}
\emph{Proof.} Assume without loss of generality that $c(u)=c(v)=1$.
Consider the restriction of $c$ to $G\setminus u$.  Every b-vertex $z$
of color~$i\ge 2$ in $G$ is still a b-vertex in $G\setminus u$,
because it cannot be that $u$ is the only neighbour of $z$ of
color~$1$.  Moreover, it cannot be that $u$ is the only b-vertex of
$G$ of color~$1$, because if it is a b-vertex then $v$ is also a
b-vertex.  But then $b(G\setminus u)\ge b(G)>\chi(G)\ge \chi(G
\setminus u)$, so $G\setminus u$ is b-imperfect, a contradiction.
Thus $c(u)\neq c(v)$.  This implies that $u$ cannot be a b-vertex,
because it has no neighbour of color~$c(v)$.  In particular, if
$N(u)=N(v)$, then the preceding argument works both ways, which leads
to the desired conclusion.  $\Box$

%%% LEMMA ABOUT TRUE TWINS
%%%
\begin{lemma}
\label{lem:ttwins} Let $G$ be a minimally b-imperfect graph, and let
$u,v$ be two true twins of $G$.  Then in any b-coloring of $G$, $u$ is
a b-vertex if and only if $v$ is.
\end{lemma}
\emph{Proof.} Let $c$ be any b-coloring of $G$ with $k$ colors.  If
$u$ is a b-vertex for $c$, then $k-2$ colors appear in $N(u)-\{v\}$.
Since $N(u) -\{u,v\} = N(v) -\{u,v\}$, $k-1$ colors (including the
color of $u$) appear in $N(v)$.  Thus, $v$ is a b-vertex.  $\Box$

%%% LEMMA ABOUT G BEING CONNECTED
%%%
\begin{lemma}[\cite{HLM}]\label{lem:conx}
Let $G$ be a minimally b-imperfect ${\cal{F}}$-free graph.  Then $G$ is
connected.
\end{lemma}
\emph{Proof.} Suppose that $G$ has several components $G_1, \ldots,
G_p$, $p\ge 2$.  By Lemma~\ref{lem:nok}, each $G_i$ has a subset $S_i$
of three vertices that induce a chordless path.  Then $G$ is
$P_4$-free, for otherwise, since a $P_4$ is in one component of $G$,
$G$ contains an $F_2$.  But then Theorem~\ref{thm:bip} is
contradicted.  Thus the lemma holds.  $\Box$

%%%% LEMMA ABOUT INDISPENDABLE VERTEX
%%%
\begin{lemma}\label{lem:map}
Let $G$ be a minimally b-imperfect graph, and let $c$ be a b-coloring
of $G$ with $b(G)$ colors.  If a vertex $x$ is not a b-vertex, then
there is a color~$f(x)$ such that $x$ is the only neighbor of
color~$c(x)$ of every $b$-vertex of color~$f(x)$.
\end{lemma}
{\it Proof.} The definition of $G$ implies $b(G) > \chi(G) \geq
\chi(G-x) = b(G-x)$. If the lemma is false, then the coloring $c$
restricted to $G-x$ still has all $b$-vertices of all colors from
$1$ to $b(G)$, implying $b(G-x) = b(G)$, a contradiction. $\Box$

\begin{lemma}\label{cl:2bv}%%%
Suppose that a graph $G$ has a coloring (not necessarily a b-coloring)
with $k$ colors where $k>\omega(G)$.  Then: \\
- If there is a b-vertex, then $G$ contains a $P_3$. \\
- If there are two b-vertices of different colors, then $G$ contains
as an induced subgraph a $P_4$, a $2P_3$ or a diamond.
\end{lemma}
{\it Proof.} To prove the first part, suppose on the contrary that $G$
contains no $P_3$.  Then every component of $G$ is a clique.  Let
$G_0$ be a component that contains a b-vertex.  Then $G_0$ has
vertices of all colors, so $G$ has a clique of size $k$, a
contradiction.

To prove the second part, let $x_i$ be a b-vertex of color~$i$ for
each $i=1, 2$.  Let $Q=N(x_1)\cap N(x_2)$.

First assume that $x_1$ sees $ x_2$.  If $Q$ contains non-adjacent
vertices $u, v$, then $x_1, x_2, u, v$ induce a diamond.  So, $Q$ is a
clique.  Then $Q\cup \{x_1, x_2\}$ is a clique, and since
$k>\omega(G)$, there is a color~$j$ that does not appear in $Q\cup
\{x_1, x_2\}$.  Since $x_1$ is a b-vertex, it has a neighbor $y$ of
color~$j$, and similarly $x_2$ has a neighbor $z$ of color~$j$.  By
the definition of $j$, vertices $y,z$ are not in $Q$, so $y$ misses
$x_2$, $z$ misses $x_1$, and $y\neq z$.  Then $y$, $x_1$, $x_2$, $z$
induce $P_4$ in $G$.

Now, we know $x_1$ misses $x_2$.  Suppose that $Q$ contains
non-adjacent vertices $u, v$.  Since $x_1$ is a b-vertex, it has a
neighbor $y$ of color~$2$.  Thus $y$ misses $x_2$ and is not in $Q$.
If $y$ misses $u$, then $y$, $x_1$, $u$, $x_2$ induce a $P_4$.  So,
$y$ sees $u$, and similarly $v$.  Then $y, x_1, u, v$ induce a
diamond.  So, $Q$ is a (possibly empty) clique.  By
Lemma~\ref{lem:simp}, for $i=1, 2$, the vertex $x_i$ is not
simplicial, so it has non-adjacent neighbors $u_i, v_i$.  If $x_1,
x_2$ are not in the same component of $G$, then $x_1, u_1, v_1, x_2,
u_2, v_2$ induce a $2P_3$.  Thus, $x_1, x_2$ are in the same component
of $G$, and so there is a chordless path between them.  If this path
has length at least three, then $G$ contains a $P_4$.  So let this
path be $x_1$-$z$-$x_2$.  We may assume that $z\neq u_1$ (or else,
symmetrically, $z\neq v_1$).  If $u_1$ misses $z$, then either $u_1$
sees $x_2$, and then $u_1, z$ are non-adjacent members of $Q$, a
contradiction, or $u_1$ misses $x_2$, and then $u_1$-$x_1$-$z$-$x_2$
is a $P_4$.  Thus we may assume that $u_1$ sees $z$.  It follows that
$z\neq v_1$, which restores the symmetry between $u_1$ and $v_1$, and
so $v_1$ too sees $z$, and $u_1, x_1, v_1, z$ induce a diamond.  Thus
the lemma holds.  $\Box$

\begin{lemma}\label{lem:xy}
Let $G$ be a $P_4$-free graph with $V(G)=X\cup Y$, where $X=\{x_1,$ $
\ldots, x_{j-1}\}$ and $Y=\{y_2, \ldots, y_j\}$ are cliques of size
$j-1\ge 2$, vertices $x_1$ and $y_j$ are adjacent, and for each $i=2,
\ldots, j-1$, vertices $x_i$ and $y_i$ are either different and not
adjacent or equal.  Then $G$ has a clique on $j$ vertices that
consists of $x_1, y_j$ and one of $x_i, y_i$ for each $i=2, \ldots,
j-1$.
\end{lemma}
\emph{Proof.} We prove the lemma by induction on the number $n$ of
indices $i$ such that $x_i\neq y_i$.  If $n=0$, then $G$ itself is the
desired clique.  Now let $n>0$, so $j\ge 3$ and, up to symmetry, we
may assume that $x_2\neq y_2$.  Let $G'$ be the graph obtained by
contracting $x_2$ and $y_2$, that is, replacing them by a vertex $z$
adjacent to all other vertices.  It is easy to see that $G'$ satisfies
the conditions of the lemma; in particular, $G'$ is $P_4$-free because
$G'\setminus z$ is equal to $G\setminus\{x_2, y_2\}$ and $z$ is
adjacent to every vertex of $G'\setminus z$.  Thus, by the induction
hypothesis, $G'$ contains a clique $K'=\{x_1, z, z_3, \ldots, z_{j-1},
y_j\}$, where, for $i=3, \ldots, j-1$, vertex $z_i$ is either $x_i$ or
$y_i$.  Let $K_x=(K'\setminus z)\cup\{x_2\}$ and $K_y=(K'\setminus z)
\cup\{y_2\}$.  If none of $K_x, K_y$ is a clique, then $x_2$ misses
some vertex $y_h$ of $K'$ and $y_2$ misses some vertex $x_g$ of $K'$,
but then $\{x_2, x_g, y_h, y_2\}$ induces a $P_4$ in $G$.  Thus one of
$K_x, K_y$ is the desired clique.  $\Box$

A set $H$ of vertices of $G$ is {\it homogeneous} if every vertex in
$G\setminus H$ either sees all or misses all vertices of $H$.  We say
that a homogeneous set $H$ of $G$ is \emph{proper} if $H \neq V(G)$.

%%% HOMOGENEOUS SET LEMMA
%%%

\begin{lemma}\label{lem:homo}
Let $G$ be a minimal counterexample to Conjecture~\ref{cj1}.  If $H$
is a proper homogeneous set in $G$, then $H$ is a clique or a stable
set.
\end{lemma}
{\it Proof.} We prove this lemma by induction on the size of $H$.
Suppose that $H$ is not a clique or a stable set.  Let $T$ be the set
of vertices of $G\setminus H$ that see all of $H$, and $Z$ be the set
of vertices of $G\setminus H$ that miss all of $H$.  So $H, T, Z$ form
a partition of $V(G)$.  Note that $T\cup Z\neq \emptyset$ by the
definition of a proper homogeneous set.  Now, by Lemma~\ref{lem:conx},
we have $T \not= \emptyset$.

Let $c$ be any b-coloring of $G$ with $k>\chi(G)$ colors.  We
may assume that the colors that appear in $H$ are $1, \ldots, h$ and
that those that have a b-vertex in $H$ are colors $1, \ldots, h_b$.
Clearly we have $h_b\le h$ and $\chi(H)\le h$.  We claim that
\begin{eqnarray}
       &h_b\le \chi(H).& \label{hbchi}
\end{eqnarray}
For suppose that $\chi(H)<h_b$.  Let $H'$ be the subgraph of $G$
induced by the vertices of $H$ that have colors $1, \ldots, h_b$, and
let $c'$ be the restriction of $c$ to $H'$.  Then the graph $H'$ has
strictly fewer vertices than $G$ (because $T\cup Z\neq\emptyset$), it
is ${\cal F}$-free, and $c'$ is a b-coloring of $H'$ with $h_b$
colors, where $h_b>\chi(H)\ge \chi(H')$, so $H'$ contradicts the
minimality of $G$.  Thus (\ref{hbchi}) holds.

\begin{eqnarray}
       &h_b>0.& \label{hbnot0}
\end{eqnarray}
For suppose that $h_b=0$.  Let $G'$ be the graph obtained from $G$ by
replacing $H$ with a stable set $S$ of size $h$ (so that all vertices
of $S$ see all of $T$ and none of $Z$ in $G'$).  We establish four
properties (i)--(iv) of $G'$.  \\
(i) $G'$ is ${\cal F}$-free.  For suppose that $G'$ contains a member
$F$ of ${\cal F}$.  If $F$ has three or more vertices of $S$, then
these vertices are pairwise twins in $F$; but no member of $F$ has
three pairwise twins.  So $F$ has at most two vertices of $S$.  Then,
since $H$ is not a clique, these vertices can be replaced by the same
number of non-adjacent vertices of $H$ so that we obtain a copy of $F$
that is an induced subgraph of $G$, a contradiction.  \\
(ii) Consider the coloring $c'$ of $G'$ that is obtained by setting
$c'(x)=c(x)$ for every $x\in T\cup Z$ and by giving colors $1, \ldots,
h$ to the vertices of $S$.  Then $c'$ is a b-coloring of $G'$ with $k$
colors, because every b-vertex in $G$ is still a b-vertex in $G'$.  \\
(iii) $\chi(G')\le \chi(G)$, because every coloring of $G$ with
$\chi(G)$ colors can be transformed into a $\chi(G)$ coloring of $G'$
by maintaining the color of the vertices in $V\setminus H$ and giving
to all vertices of $S$ the color of a fixed vertex of $H$.  \\
(iv) $|V(G')|\le |V(G)|$; and if $|V(G')|= |V(G)|$ then, since $H$ is
not a stable set, we have $|E(G')|< |E(G)|$.  \\
It follows from Properties (i)--(iv) that $G'$ contradicts the
minimality of $G$.  Thus (\ref{hbnot0}) holds.

Note that, since $T\neq \emptyset$, we have $k\ge h+1$.  By
(\ref{hbnot0}), all colors $h+1, \ldots, k$ appear in $T$.  We claim
that
\begin{eqnarray}
       &h>\chi(H).& \label{hchi}
\end{eqnarray}
For suppose on the contrary that $h=\chi(H)$.  Let $G'$ be the graph
obtained from $G$ by replacing $H$ with a clique $K$ of size $h$ (so
that all vertices of $K$ see all of $T$ and none of $Z$ in $G'$).  We
establish four properties (i)--(iv) of $G'$.  \\
(i) $G'$ is ${\cal F}$-free.  For suppose that $G'$ contains a member
$F$ of ${\cal F}$.  If $F$ has three or more vertices of $K$, then
these vertices are pairwise adjacent twins in $F$; but no member of
$F$ has three pairwise adjacent twins.  So $F$ has at most two
vertices of $K$.  Then these vertices can be replaced by the same
number of adjacent vertices of $H$, so that we obtain a copy of $F$
that is an induced subgraph of $G$, a contradiction.  \\
(ii) Consider the coloring $c'$ of $G'$ obtained by setting
$c'(x)=c(x)$ for every $x\in T\cup Z$ and by giving
colors $1, \ldots, h$ to the vertices of $K$.  Then $c'$ is a
b-coloring of $G'$ with $k$ colors, because every b-vertex of color
$>h$ in $G$ is still a b-vertex in $G'$, and every vertex of $K$ is a
b-vertex in $G'$ since all colors $h+1, \ldots, k$ appear in $T$.  \\
(iii) $\chi(G')\le \chi(G)$, because every coloring of $G$ with
$\chi(G)$ colors must use at least $\chi(H)=h$ colors on $H$ and can
be turned into a $\chi(G)$ coloring of $G'$ by giving colors $1,
\ldots, h$ to the vertices of $K$.  \\
(iv) Since $H$ is not a clique, we have $|H|>\chi(H)$, so $G'$ has
strictly fewer vertices than $G$.  \\
It follows from Properties (i)--(iv) that $G'$ contradicts the
minimality of $G$.  Thus (\ref{hchi}) holds.

Now, we can apply the first part of Lemma~\ref{cl:2bv} to $H$ and to
the restriction of $c$ to $H$, which implies that $H$ contains
a $P_3$.  Note that (\ref{hbchi}) and (\ref{hchi}) imply $h_b<h$.

%%%%%

Now we claim that:
\begin{eqnarray}
       &\mbox{$H$ contains no $P_4$ and no $2P_3$.}& \label{hfree}
\end{eqnarray}
For suppose that $H$ contains a $P_4$ or a $2P_3$, with vertex-set
$X$.  We distinguish between two cases.  \\
{\it Case 1: $h_b=1$.} Let $x$ be a b-vertex of $H$ with $c(x)=1$.
Let $G'$ be the graph obtained from $G$ by removing every edge whose
two endvertices are in $H\setminus \{x\}$.  We establish three
properties (i)--(iii) of $G'$.  \\
(i) $G'$ is ${\cal F}$-free.  For suppose that $G'$ contains a member
$F$ of ${\cal F}$.  Then $F\cap H$ is a homogeneous set of $F$; and
since $H$ (in $G'$) contains no $P_4$, no $2P_3$ and no diamond, this
is possible only if $F\cap H$ either is a $P_3$ or has at most two
vertices; and in either case it is possible to replace $F\cap H$ by a
subgraph of $H$ in $G$ that is isomorphic to $F\cap H$, so that we
obtain a copy of $F$ that is an induced subgraph of $G$, a
contradiction.  \\
(ii) $c$ is a b-coloring of $G'$ with $k$ colors (because every
b-vertex in $G$ is still a b-vertex in $G'$).  \\
(iii) $\chi(G')\le \chi(G)$, clearly.  \\
It follows from Properties (i)--(iii) that if $G'$ has strictly fewer
edges than $G$, then $G'$ contradicts the minimality of $G$.  So it
must be that $G'=G$.  This means that every edge in $H$ is adjacent to
$x$.  Thus $H$ contains no $P_4$ and no $2P_3$ as desired.  %
\\
{\it Case 2: $h_b\ge 2$.} So $h\ge 3$.  Let $u$ be any b-vertex of
color~$h$.  Since $h_b<h$, vertex $u$ is not in $H$, and since
color~$h$ appears in $H$ it is not in $T$, so we have $u\in Z$.  Note
that $Z$ contains no $P_3$, for otherwise, if $X'$ is the vertex-set
of a $P_3$ in $Z$, then $X\cup X'$ induces an $F_2$ or $F_3$.
Therefore every component of $Z$ is a clique.  Let $Y$ be the
component of $Z$ that contains $u$.  Since $u$ is a b-vertex, all
colors $1, \ldots, h$ must appear in $Y$.  So $|Y|\ge h\ge 3$.
Suppose that $Y$ is not homogeneous.  Then there are vertices $y_1,
y_2\in Y$ and a vertex $t$ that sees $y_1$ and misses $y_2$.  Clearly
$t\in T$.  Let $y_3$ be a vertex of $Y\setminus\{y_1, y_2\}$.  If $t$
misses $y_3$, then $X\cup\{t, y_1, y_2, y_3\}$ induces an $F_8$ or
$F_9$.  If $t$ sees $y_3$ then, letting $X''$ be the vertex-set of a
$P_3$ in $H$, we obtain that $X''\cup\{t, y_1, y_2, y_3\}$ induces an
$F_5$.  Therefore $Y$ is a homogeneous set.  It follows that all
vertices of $Y$ are b-vertices, and so $Y$ contains b-vertices of
colors $1, \ldots, h$.  Let $G'$ be the graph obtained from $G$ by
removing every edge whose two endvertices are in $H$.  We establish four
properties (i)--(iv) of $G'$.  \\
(i) $G'$ is ${\cal F}$-free.  For suppose that $G'$ contains a member
$F$ of ${\cal F}$.  If $F$ has three or more vertices of $H$, then
these vertices are pairwise non-adjacent twins in $F$; but no member
of $F$ has three pairwise non-adjacent twins.  So $F$ has at most two
vertices of $H$.  Then these vertices can be replaced by the same
number of non-adjacent vertices of $H$ in $G$, so that we obtain a
copy of $F$ that is an induced subgraph of $G$, a contradiction.
\\
(ii) $c$ is a b-coloring of $G'$ with $k$ colors (because every
b-vertex of color~$>h$ in $G$ is still a b-vertex in $G'$, and $Y$
contains b-vertices of colors $1, \ldots, h$).  \\
(iii) $\chi(G')\le \chi(G)$, obvious.  \\
(iv) Since $H$ is not a stable set, $G'$ has strictly fewer edges than
$G$.  \\
It follows from Properties (i)--(iv) that $G'$ contradicts the
minimality of $G$.  Thus (\ref{hfree}) holds.

Suppose that $H$ has at least two components.  Since $H$ contains no
$2P_3$, one of these components $K$ is a clique.  Let $x$ be a
b-vertex in $H$.  If $x\in K$, then all colors $1, \ldots, h$ appear
in $K$ and all vertices of $K$ are b-vertices; thus $h_b=h$, a
contradiction.  So $K$ contains no b-vertex, and $x$ is in another
component of $H$.  It follows that all colors that appear in $K$ also
appear in $H\setminus K$.  Consider the graph $G\setminus K$ and the
restriction $c'$ of $c$ to that graph.  Then $c'$ is a b-coloring of
$G\setminus K$ with $k$ colors, and $k>\chi(G)\ge \chi(G\setminus K)$,
so $G\setminus K$ contradicts the minimality of $G$.  So $H$ is
connected.

Since $H$ is $P_4$-free, connected, and has more than one vertex, a
classical theorem of Seinsche \cite{sei74} states that $H$ can be
partitioned into two non-empty sets $Q, S$ such that every vertex of
$Q$ is adjacent to every vertex of $S$.  Now, each of $Q, S$ is a
homogeneous set that is strictly smaller than $H$.  By the induction
hypothesis, each of $Q, S$ is a clique or a stable set.  If $Q$ and
$S$ are two cliques, then $H$ is a clique as desired.  If $Q$ and $S$
are two stable sets (of size at least two), then
Lemma~\ref{lem:compuv} implies that no vertex of $H$ is a b-vertex,
which contradicts $h_b>0$.  Therefore we may assume up to symmetry
that $Q$ is a clique and $S$ is a stable set of size at least two.
Let $\ell = |S|$ and $S=\{s_1, \ldots, s_\ell\}$.  By
Lemma~\ref{lem:compuv}, all vertices of $S$ have different colors and
are not b-vertices.  Up to renaming colors, we will assume that $s_i$
has color $i$ for each $i=1, \ldots, \ell$.  Since $H$ is complete to
$T$, the set $T$ contains no vertex of color $1, \ldots, \ell$; thus
we know that:
\begin{equation}\label{dellinz}
\mbox{$Z$ contains b-vertices of all colors $1, \ldots, \ell$.}
\end{equation}

We claim that:
\begin{equation}\label{qonlyb}
\mbox{Each vertex of $Q$ is the only b-vertex of its color.}
\end{equation}
First note that, since $h_b>0$ and the vertices of $S$ are not
b-vertices, some vertex of $Q$ is a b-vertex; and since the vertices
of $Q$ are pairwise adjacent twins, by Lemma~\ref{lem:ttwins} they are
all b-vertices.  Now suppose that some vertex $q$ of $Q$ is not the
only b-vertex of its color, say color~$\ell+1$.  Let $G'$ be the graph
obtained from $G$ by removing every edge between $q$ and $S$.  Note
that the subgraph $G'[H]$ contains no $P_4$ or $2P_3$.  We establish
four properties (i)--(iv) of $G'$.  \\
(i) $G'$ is ${\cal F}$-free.  For suppose that $G'$ contains a member
$F$ of ${\cal F}$.  If $F$ has at most two vertices of $H$, then these
vertices can be replaced by the same number of vertices of $H$ in $G$,
so that we obtain a copy of $F$ that is an induced subgraph of $G$, a
contradiction.  If $F$ has three or more vertices of $H$, then, since
$G'[H]$ contains no $P_4$ or $2P_3$, it must be (by examination of the
list ${\cal F}$) that $F\cap H$ is either a $P_3$ or diamond; but this
implies that there is a $P_3$ or diamond in $H$, and so, $G$ contains
a copy of $F$.  \\
(ii) $c$ is a b-coloring of $G'$ with $k$ colors, because every
b-vertex of color $\neq\ell+1$ in $G$ is still a b-vertex in $G'$ and
there is a b-vertex of color~$\ell+1$ different from $q$.  \\
(iii) $\chi(G')\le \chi(G)$, clearly.  \\
(iv) $G'$ has strictly fewer edges than $G$, clearly.  \\
It follows from Properties (i)--(iv) that $G'$ contradicts the
minimality of $G$.  Thus (\ref{qonlyb}) holds.

Pick a vertex $q\in Q$, and assume that its color is $\ell+1$.  We
note that:
\begin{equation}\label{znop4}
\mbox{$Z$ contains no $P_4$ and no $2P_3$.}
\end{equation}
For otherwise the union of the vertices of such a subgraph of $Z$
with vertices $q, s_1, s_2$ induces an $F_2$ or $F_3$.  Thus
(\ref{znop4}) holds.

\begin{equation}\begin{minipage}{0.85\linewidth}\label{cplus}
Let $C$ be a clique in $Z$  and let $i$ be a color not in $C$ such
that every vertex of $C$ has a neighbor of color $i$.  Then there is a
vertex of color $i$ that is adjacent to all of $C$.
\end{minipage}
\end{equation}
Pick a vertex $u$ of color $i$ that has the most neighbors in $C$.
Suppose that $u$ has a non-neighbor $y$ in $C$.  By the hypothesis we
know that $y$ has a neighbor $v$ of color $i$.  By the choice of $u$,
there exists a vertex $x$ of $C$ that sees $u$ and misses $v$.  So
$\{u, v, x, y\}$ induce a $P_4$.  By (\ref{znop4}), one of $u, v$ is
not in $Z$ and thus is in $T$.  If exactly one of $u,v$ is in $T$,
then $\{u, v, x, y, q\}$ induces an $F_1$; if both $u,v$ are in $T$,
then $\{u, v, x, y, q, s_1\}$ induces an $F_{16}$, a contradiction.
So $u$ is adjacent to all of $C$.  Thus (\ref{cplus}) holds.

\begin{equation}\begin{minipage}{0.85\linewidth}\label{zkl}
For every set $J\subseteq\{1, \ldots, \ell\}$, $Z$ contains a clique
of b-vertices of all colors from $J$.
\end{minipage}
\end{equation}
We prove (\ref{zkl}) by induction on $|J|$.  The assertion holds when
$|J|=1$ by (\ref{dellinz}).  Let us now assume that $|J|\ge 2$.  To
simplify notation put $J=\{1, 2, \ldots, j\}$.  By the induction
hypothesis, $Z$ contains a clique $X=\{x_1, \ldots, x_{j-1}\}$ where
each $x_i\in X$ is a b-vertex of color $i$.  For each $i=1, \ldots,
j-1$, vertex $x_i$ has a neighbor of color $j$, so, by (\ref{cplus}),
there exists a vertex $u_j$ of color $j$ that is adjacent to all of
$X$; moreover $u_j$ is in $Z$ since color $j$ does not appear in $T$.
If $u_j$ is a b-vertex, then the desired conclusion holds with clique
$X\cup\{u_j\}$.  So let us assume that there is a color $g\neq j$ such
that $u_j$ has no neighbor of color $g$.  Clearly $g\notin\{1, \ldots,
j-1\}$.  Every member of $X$ has a neighbor of color $g$, so, by
(\ref{cplus}), there is a vertex $v_g$ of color $g$ that is adjacent
to all of $X$.  Note that $v_g$ is in $Z\cup T$.  Similarly, by the
induction hypothesis, $Z$ contains a clique $Y=\{y_2, \ldots, y_j\}$
where each $y_i\in Y$ is a b-vertex of color $i$.  By the same
argument as for $X$, there exists in $Z$ a vertex $u_1$ that is
adjacent to all of $Y$, vertex $u_1$ is not a b-vertex, so there is a
color $h\notin\{1, \ldots, j\}$ such that $u_1$ has no neighbor of
color $h$, and there is a vertex $v_h$ of color $h$ in $Z\cup T$ that
is adjacent to all of $Y$.  \\
If $x_1$ sees $y_j$, then $X$ and $Y$ satisfy the hypothesis of
Lemma~\ref{lem:xy}, so there exists a clique that contains $x_1, y_j$
and one of $x_i, y_i$ for each $i=2, \ldots, j-1$, which is the
desired clique for (\ref{zkl}).  Let us now assume that $x_1$ misses
$y_j$.  Then $u_1$ misses $u_j$, for otherwise $\{u_1, x_1, u_j ,
y_j\}$ induces a $P_4$ in $Z$, which contradicts (\ref{znop4}).  We
have $v_g\neq v_h$, for otherwise $\{u_j, x_1, u_1, y_j, v_g\}$
induces an $F_1$.  Suppose that $v_g, v_h$ are both in $Z$.  Then
$v_g$ misses $u_1$, for otherwise $\{u_j, x_1, v_g, u_1\}$ induces a
$P_4$ in $Z$.  Similarly, $v_g$ misses $y_j$, and $v_h$ misses both
$u_j, x_1$.  Then $v_g$ misses $v_h$, for otherwise $\{x_1, v_g, y_j,
v_h\}$ induces a $P_4$ in $Z$.  But then $\{u_j, x_1, v_g, u_1, y_j,
v_h\}$ induces a $2P_3$ in $Z$, which contradicts (\ref{znop4}).
Therefore we may assume up to symmetry that $v_g$ is in $T$.  If $j\ge
3$, then $\{s_1, s_2, q, v_g, x_1, x_2, u_j\}$ induces an $F_5$.  So
$j=2$.  \\
Suppose that $u_1, u_2$ are in the same component of $Z$.  So there is
a chordless path in $Z$ between them, and since $Z$ contains no $P_4$,
there is a vertex $z\in Z$ that sees both $u_1, u_2$.  Then $z$ sees
$x_1$, for otherwise $\{x_1, u_2,$ $z, u_1\}$ induces a $P_4$, and
similarly $z$ sees $y_2$.  Then $v_g$ misses $z$, for otherwise $\{q,
s_1, s_2, v_g, z, x_1, u_2\}$ induces an $F_5$.  Then $v_g$ sees
$u_1$, for otherwise $\{q, v_g, x_1, z, u_1\}$ induces an $F_1$, and
similarly $v_g$ sees $y_2$.  But then $\{q, s_1, s_2, v_g, u_1, y_2,
z\}$ induces an $F_5$.  Therefore $u_1$ and $u_2$ are in different
components of $Z$.  \\
Let $X'$ be the component of $Z$ that contains $x_1$ and $u_2$,
and let $Y'$ be the component that contains $y_2$ and $u_1$.  If
both $X'$ and $Y'$ contain a $P_3$, then each of $H, X', Y'$
contains a $P_3$, so $G$ contains an $F_3$, a contradiction.  Thus
one of $X', Y'$ is a clique.\\
Since $x_1$ is a b-vertex, it has a neighbor $r$ of color $\ell+1$
(the color of $q$), and this neighbor is in $X'$ because color
$\ell+1$ does not appear in $T$. Note that if $v_g$ has a neighbor
in $Y'$, then it is $Y'$-complete, for otherwise $\{u_2, x_1,
v_g,$ $ z, z'\}$ induces an $F_1$ for some
adjacent vertices $z, z'\in Y'$.  \\
Suppose that $Y'$ is not a clique (and thus $X'$ is a clique); so
it contains a $P_3$ with vertices $y, y', y''$.  Then $v_g$ sees
at least one of $y, y', y''$, for otherwise $\{q, v_g, x_1, u_2,$
$ y, y', y''\}$ induces an $F_2$; and so $v_g$ is $Y'$-complete.
Now, $v_g$ misses $r$, for otherwise $\{s_1, q, s_2, v_g,$ $ x_1,
r, u_2\}$ induces an $F_5$. But then $\{s_1, q, s_2, v_g, y, y',
y'', x_1, u_2, r\}$ induces an $F_9$.  Therefore $Y'$ is a clique,
and thus $v_h \in T$. It follows by symmetry that $X'$ is also a
clique, and if $v_h$ has a neighbor in $X'$ then it is
$X'$-complete. Suppose that $v_g$ and $v_h$ are not adjacent. If
$v_h$ has no neighbor in $X'$, then $\{u_2, x_1, v_g, q, v_h\}$
induces an $F_1$.  So $v_h$ is $X'$-complete, and similarly $v_g$
is $Y'$-complete.  But then $\{u_1, u_2, v_g, v_h, q\}$ induces an
$F_1$.  Therefore $v_g$ and $v_h$ are adjacent.
Suppose $v_g$ is $Y'$-complete. If $v_h$ is $X'$-anti-complete,
then $\{q, v_h, y_2, u_1, v_g, x_1, u_2, r\}$ induces an $F_8$
(note that $v_g$ misses $r$, for otherwise $\{s_1, q, s_2, v_g,
x_1, r, u_2\}$ induces an $F_5$.). If $v_h$ is $X'$-complete, then
$\{u_2, x_1, v_g,$ $ v_h, y_2, u_1\}$ induces a $F_4$.  So $v_g$
has no neighbor in $X'$, and similarly $v_h$ has no neighbor in
$Y'$. But then $\{u_2, x_1, v_g, v_h, y_2\}$ induces an $F_1$, a
contradiction. Thus (\ref{zkl}) holds.

By (\ref{zkl}), we know that $Z$ contains a clique of b-vertices
$b_1, \ldots, b_\ell$, which we call $D_\ell$.  By (\ref{cplus})
there is a vertex $u_{\ell+1}$ of color ${\ell+1}$ (the color of
$q$) that is adjacent to all of $D_\ell$, and $u_{\ell+1}$ is in
$Z$ because color ${\ell+1}$ does not appear in $T$.  By
(\ref{qonlyb}), $q$ is the only b-vertex of color $\ell+1$, so
there exists a color $m$ such that $u_{\ell+1}$ has no neighbor of
color $m$.  Clearly, $m>\ell$.  By (\ref{cplus}) there is a vertex
$u_m$ of color $m$ that is adjacent to all of $D_{\ell}$. If $u_m$
is in $T$, then $\{q, s_1, s_2, u_m, b_1, b_2, u_{\ell+1}\}$
induces an $F_5$.  So $u_m$ is in $Z$.  If $|Q|\ge 2$, pick any
vertex $q'\in Q$ with $q'\neq q$; then $\{q, q', s_1, s_2, b_1,
b_2, u_{\ell+1}, u_m\}$ induces an $F_6$.  So $Q=\{q\}$. Since $q$
has a neighbor of color $m$, and $m>\ell$, there is a vertex $t_m$
of color $m$ in $T$.  Then $t_m$ misses one of $b_1, b_2$, for
otherwise $\{q, s_1, s_2, t_m, b_1, b_2, u_m\}$ induces an $F_5$.
Say $t_m$ misses $b_1$.  Recall that $u_{\ell+1}$ has no neighbor
of color $m$.  Then $\{q, s_1, s_2, t_m, b_1, b_2, u_{\ell+1},
u_m\}$ induces an $F_6$ (if $t_m$ misses $b_2$) or $F_7$ (if $t_m$
sees $b_2$), a contradiction.  This concludes the proof of
Lemma~\ref{lem:homo}. $\Box$

%%% END NEW homogeneous SET LEMMA

%%%
%%%
%%% SUBSECTION ABOUT THE CASE G CONTAINS A C5
%%%
%
\section{Graphs that contain a $C_5$}
The following lemma was proved in \cite{HLM}.
%
% LEMMA ABOUT C5
%
\begin{lemma}[\cite{HLM}]\label{lem:c5}
Let $G$ be an ${\cal F}$-free graph that contains a $C_5$.  Then
$V(G)$ can be partitioned into sets $X_1, \ldots, X_6, T, Z$ such
that:
\begin{enumerate}
   \item
Each of $X_1, \ldots, X_5$ is not empty.
\item
For every $j$ modulo $5$, $X_j$ is complete to $X_{j+1}$.
\item
For every $j$ modulo $5$ and $j\neq 4$, $X_j$ is anti-complete to
$X_{j+2}$, and some vertex of $X_1$ misses a vertex of $X_4$.
\item
$X_6$ is complete to $X_2\cup X_3\cup X_5$ and anti-complete to
$X_1\cup X_4$.
\item
$X_2, X_3, X_5$ are stable sets.
\item
The sets $X'_1=\{x\in X_1\mid x \mbox{ has a non-neighbour in } X_4\}$
and $X'_4=\{x\in X_4\mid x \mbox{ has a non-neighbour in } X_1\}$ are
stable sets, and there is no edge between $X'_1$ and $X_1\setminus
X'_1$ and no edge between $X'_4$ and $X_4\setminus X'_4$.
  \item
At least one of $X_1\setminus X'_1, X_4\setminus X'_4, X_6$ is empty.
\item
Any two non-adjacent vertices of $X_1$ have inclusionwise comparable
neighbourhoods in $V(G)\setminus X_1$, and the same holds for $X_4$
and $X_6$.
\item
$T$ is complete to $X_1\cup \cdots \cup X_6$.
\item
$Z$ is anti-complete to $X'_1\cup X_2\cup X_3\cup X'_4\cup X_5$; and
if $X_6\neq\emptyset$, then $Z$ is anti-complete to $X_1\cup X_2\cup
X_3\cup X_4\cup X_5$.
\item
Every component of $Z$ is a clique and is a homogeneous set in
$G\setminus T$.
\end{enumerate}
\end{lemma}
%
%%%%% THEOREM about C5
%
\begin{theorem}\label{thm:noc5}
Let $G$ be a minimal counterexample to Conjecture~\ref{cj1}.  Then $G$
contains no $C_5$.
\end{theorem}
\emph{Proof.} Suppose on the contrary that $G$ contains a $C_5$.  So
$\chi(G) \geq 3$ and consequently $b(G) \geq 4$.  Let $c$ be a
b-coloring of $G$ with $b(G)$ colors.  For each $i\in\{1, \ldots,
b(G)\}$, let $b_i$ be a b-vertex of color~$i$.  Since $G$ contains a
$C_5$, it has a partition into sets $X_1, \ldots, X_6, T, Z$ with the
notation and properties given in Lemma~\ref{lem:c5}.  For each
$j\in\{2, 3, 5\}$ let $a_j$ be an arbitrary vertex in $X_j$, and let
$a_1, a_4$ be non-adjacent vertices of $X_1$ and $X_4$ respectively.
Such vertices exist by item~3 of Lemma~\ref{lem:c5}.  For each
$j\in\{1, 4\}$, let $X''_j=X_j\setminus X'_j$.  So every vertex of
$X''_j$ is $X_{5-j}$-complete.  Lemmas~\ref{lem:c5}
and~\ref{lem:compuv} imply easily the following two facts:
\begin{equation}\begin{minipage}{0.85\linewidth}\label{eq:tw235}
For each $j\in\{2, 3, 5\}$, any two vertices in $X_j$ are twins and
have different colors.  If $X_j$ contains a b-vertex, then $|X_j|=1$.
\end{minipage}\end{equation}
\begin{equation}\begin{minipage}{0.85\linewidth}\label{eq:2xi}
For each $j\in\{1, 4\}$, if $u, v\in X'_j$, then either $N(u)\subseteq
N(v)$ or $N(v)\subseteq N(u)$.  If $u\in X'_j$ and $v\in X''_j$, then
$N(u)\subset N(v)$.  If $X'_j$ contains a b-vertex $u$, then
$X''_j=\emptyset$ and any vertex $v$ of $X'_j\setminus \{u\}$
satisfies $N(v)\subset N(u)$ and is not a b-vertex.
\end{minipage}\end{equation}
Let $X'=X'_1\cup X_2\cup X_3\cup X'_4\cup X_5$.  Note that $X'$
contains $a_1, \ldots, a_5$, which induce a $C_5$; and that every
vertex of $Z$ misses every vertex of $X'$, by Lemma~\ref{lem:c5}.

\begin{equation}\begin{minipage}{0.85\linewidth}\label{eqy3}
If a component $Y$ of $Z$ has at least three vertices, then $Y$ is a
homogeneous set and a clique.
\end{minipage}\end{equation}
Recall that $Y$ is a clique by item~11 of Lemma~\ref{lem:c5}.  Suppose
that $Y$ is not homogeneous.  Then, by item~11 again and by the
definition of $Z$, there is a vertex $t\in T$ and vertices $y_1,
y_2\in Y$ such that $t$ sees $y_1$ and misses $y_2$.  Consider a
vertex $y_3\in Y\setminus\{y_1, y_2\}$.  If $t$ sees $y_3$, then $\{t,
y_1, y_2, y_3, a_1, a_2, a_3\}$ induces an $F_5$; if $t$ misses $y_3$,
then $\{t, y_1, y_2, y_3, a_1, a_2, a_3, a_4\}$ induces an $F_8$, a
contradiction, so (\ref{eqy3}) holds.

\begin{equation}\begin{minipage}{0.85\linewidth}\label{eqzb}
If a component $Y$ of $Z$ contains a b-vertex, then $Y$ is a
homogeneous set and a clique, $|Y| \ge 3$, every vertex of $Y$ is a
b-vertex, and every color that appears in $X'$ appears in~$Y$.
\end{minipage}\end{equation}
Let $M=N(Y)$.  By Lemma~\ref{lem:c5} we have $M\subseteq X''_1\cup
X''_4\cup X_6\cup T$.  Since $Y$ contains a b-vertex, all colors
appear in $Y\cup M$.  \\
Suppose that $M\cap X_6\neq \emptyset$.  So there is an edge $yu_6$
with $y\in Y$ and $u_6\in X_6$.  Then there is no edge $x_1x_4$ with
$x_1\in X_1$ and $x_4\in X_4$, for otherwise, by item~10 of
Lemma~\ref{lem:c5}, $\{y, u_6, a_2, x_1, x_4\}$ induces an $F_1$.  It
follows that $X_1=X'_1$ and $X_4=X'_4$, and so these are stable sets.
Now, by items~4 and~10 of Lemma~\ref{lem:c5}, every vertex $x\in
X_1\cup X_4$ satisfies $N(x)\subseteq N(x_6)$ for every $x_6\in X_6$,
and so, by Lemma~\ref{lem:compuv}, the color of $x$ is different from
all colors in $X_6$.  Clearly the color of every vertex in $X_2\cup
X_3\cup X_5$ is different from all colors in $X_6$; and the color of
every vertex in $X'$ is also different from all colors in $T$.  Thus
every color that appears in $X'$ does not appear in $M$, and
consequently it appears in $Y$.  So $|Y|\ge 3$, since at least three
colors appear in $X'$ because $X'$ contains a $C_5$.  By (\ref{eqy3}),
$Y$ is a homogeneous set and a clique, and so every vertex of $Y$ is a
b-vertex.  Thus in this case (\ref{eqzb}) holds.  \\
Now suppose that $M\cap (X_1\cup X_4) \neq\emptyset$.  So, up to
symmetry, there exist adjacent vertices $y\in Y$ and $x_1\in X_1$.  By
item~10 of Lemma~\ref{lem:c5}, $x_1$ sees all of $X_4$ (that is,
$x_1\in X''_1$), and $X_6$ is empty.  Now every vertex $x$ in
$X'_1\cup X_3$ satisfies $N(x) \subseteq N(v_1)$ for every $v_1\in
X''_1$, and so, by Lemma~\ref{lem:compuv}, the color of $x$ is
different from all colors in $X''_1$.  Clearly the color of every
vertex in $X_2\cup X_4\cup X_5$ is different from all colors in
$X''_1$; and the color of every vertex in $X'$ is also different from
all colors in $T$.  Thus every color that appears in $X'$ does not
appear in $M$, and consequently it appears in $Y$.  As above, this
implies that $|Y|\ge 3$, $Y$ is a homogeneous set and a clique, and
every vertex of $Y$ is a b-vertex.  Thus in this case too (\ref{eqzb})
holds.  \\
Finally suppose that $M\subset T$.  Then every vertex of $M$ is
adjacent to every vertex of $X'$, so we conclude immediately as
above. Thus (\ref{eqzb}) holds.

\begin{equation}
 \begin{minipage}{0.8\linewidth}\label{threevertices}
If some component $Y$ of $Z$ contains a b-vertex, then there does
not exist vertices $a \in X_3 \cup X_4, b_h \in T \cup X''_4$ such
that $c(b_h)=h$, $a$ is not a b-vertex but $b_h$ is, and $f(a)=h$,
that is, every b-vertex of color $h$ has $a$ as its unique
neighbor of color $c(a)$.
 \end{minipage}
\end{equation}
%
%\marginpar{see comment file}
%
Suppose $Y, a , b_h$ exist. Let $y, y'$ be two vertices of $Y$.
For simplicity, assume $c(a)=1$. So $Y$ satisfies the properties
given in (\ref{eqzb}). By (\ref{eqzb}), we may assume that $b_1\in
Y$. Let $M=N(Y)$. By Lemma~\ref{lem:c5} we have $M\subseteq
X''_1\cup X''_4\cup X_6\cup T$. By (\ref{eqzb}), the vertices of
$Y$ must have a neighbor $u_h$ of color $h$, so $u_h\in M$. If
$u_h \in X_6$, then $b_h \in X''_4$, and so $b_1, u_h, a_3, b_h,
a_1$ induces an $F_1$, a contradiction. If $u_h \in X_1$, then it
sees $b_h$, a contradiction. So $u_h$ is in $T\cup X''_4$. In
other words, $b_h$ and $u_h$ are in the same co-component $K$ of
$T\cup X''_4$ (with either $K\subset T$ or $K\subseteq X''_4$).
Since $u_h$ is adjacent to $b_1$, it is not a b-vertex, and so it
has no neighbor of color~$i$ for some $i\neq h$. Since $u_h$ is
$Y$-complete, color~$i$ does not appear in $Y$. The vertices of
$Y$ must have a neighbor $u_i$ of color $i$, so $u_i\in M$. If
$u_i \in X_6$, then $u_h, b_h \in X''_4$, and so $b_1, u_i, a_3,
b_h, a_1$ induce an $F_1$. It follows that $u_i \in K$. Then $b_h$
is not adjacent to $u_i$, for otherwise $\{a_3, a_5, b_h, u_i,
u_h, y, y'\}$ induces an $F_{11}$.  Vertex $b_h$ must have a
neighbor $v_i$ of color $i$. If $v_i$ is in $M$, then it can play
the role of $u_i$ (in particular, $v_i \not\in X_1$ and $v_i$ sees
$x_3, x_5$) and we find an $F_{11}$ again. So $v_i\notin M$. If
$v_i$ is in $Z$, then $\{v_i, b_h, a_3, u_i, y\}$ induces an
$F_1$.  So $v_i\notin Z$. If $v_i$ is in $X_1\cup X_2\cup X_3\cup
X_5$, then it is either $K$-complete or $K$-anticomplete, which is
impossible (as $v_i$ sees $b_h$ and misses $u_i$).  So $v_i$ is in
$T\cup X''_4$ (that is, in $K$). But then $\{a_3, a_5, b_h, u_i,
u_h, v_i, y, y'\}$ induces an $F_{13}$. We have established
(\ref{threevertices}).

\begin{equation}
     \mbox{$Z$ contains no b-vertex.}\label{znob}
\end{equation}
For suppose that $Y$ is a component of $Z$ that contains a b-vertex.
So $Y$ satisfies the properties given in (\ref{eqzb}).  Let $M=N(Y)$.
By Lemma~\ref{lem:c5} we have $M\subseteq X''_1\cup X''_4\cup X_6\cup
T$.  There must be an edge $xz$ with $x\in X_1\cup\cdots\cup X_6$ and
$z\in Z$, for otherwise $X_1\cup\cdots\cup X_6$ is a homogeneous set,
which contradicts Lemma~\ref{lem:homo}.  By item~10 of
Lemma~\ref{lem:c5}, $x$ is in $X''_1\cup X''_4\cup X_6$.  Up to
symmetry we distinguish two cases.  \\
Case 1: $x\in X''_1$.  Then item~10 of Lemma~\ref{lem:c5} implies
$X_6=\emptyset$.  So $M\subseteq X''_1\cup X''_4\cup T$.  Recall
from item~9 of Lemma~\ref{lem:c5} and the definition of $X''_1,
X''_4$ that the three sets $X''_1, X''_4$ and $T$ are complete to
each other. Since $X_6=\emptyset$, we have $N(a_3)\subset N(x)$,
so by Lemma~\ref{lem:compuv}, $a_3$ is not a b-vertex and we can
apply Lemma~\ref{lem:map}.  Let $c(a_3)=1$ and $f(a_3)=h>1$.  By
(\ref{eqzb}), we may assume that $b_1\in Y$.  The definition of
$f(a_3)$ means that every b-vertex of color $h$ is in $N(a_3)$ and
not in $N(b_1)$.  Recall that $N(a_3)=X_2\cup X_4\cup T$. Moreover
$b_h$ is not in $X_2\cup X'_4$, for otherwise, by (\ref{eqzb})
there would be a b-vertex of color $h$ in $Y$, a contradiction.
Thus $b_h$ is in $T\cup X''_4$. But, the existence of $a_3$ and
$b_h$ is contradicted by (\ref{threevertices}).

Case 2: $x\in X_6$.  Then item~10 of Lemma~\ref{lem:c5} implies
that there is no edge between $Z$ and $X_1\cup X_4$.  So
$M\subseteq X_6\cup T$.  By item~9 of Lemma~\ref{lem:c5}, $X_6$
and $T$ are complete to each other.  If there is an edge between
two vertices $x_1 \in X_1$ and $x_4 \in X_4$, then the five
vertices $z, x, a_3, x_4, x_1$ induce an $F_1$.  Item~6 of
Lemma~\ref{lem:c5} implies $X_4$ is a stable set. Pick any $a\in
X_4$; so $a$ has no neighbor in $X_1$. We have $N(a)\subset N(x)$,
so by Lemma~\ref{lem:compuv}, $a$ is not a b-vertex and we can
apply Lemma~\ref{lem:map}.  Let $c(a)=1$ and $f(a)=h>1$.  By
(\ref{eqzb}), we may assume that $b_1\in Y$. The definition of
$f(a)$ means that every b-vertex of color $h$ is in $N(a)$ and not
in $N(b_1)$.  Recall that $N(a)=X_3\cup X_5\cup T$. Moreover,
$b_h$ is not in $X_3\cup X_5$, for otherwise, by (\ref{eqzb})
there would be a b-vertex of color $h$ in $Y$, a contradiction.
Thus, $b_h$ is in $T$. But, the existence of $a$ and $b_h$ is
contradicted by (\ref{threevertices}).

%%%   CLAIM X6 IS EMPTY

\begin{equation}
X_6=\emptyset.\label{x60}
\end{equation}
For suppose the contrary.  Pick any $x_6\in X_6$.  By item~10 of
Lemma~\ref{lem:c5}, there is no edge between $Z$ and
$X_1\cup\cdots\cup X_5$.  By item~7, we may assume that $X''_1=
\emptyset$; so $X_1$ is a stable set.  Suppose that every vertex of
$X'_4$ has a neighbor in $X_1$.  Consider a vertex $u$ of $X_1$ for
which $N(u)\cap X'_4$ is maximal; then, the first line of
(\ref{eq:2xi}) implies $X'_4\subset N(u)$, and so $u\in X''_1$, a
contradiction.  Therefore some vertex $a\in X_4$ has no neighbor in
$X_1$.  Let $c(a)=1$.  We have $N(a)\subset N(x)$ for every $x\in
X_6\cup X_4\setminus\{a\}$, so Lemma~\ref{lem:compuv} implies that
color~$1$ does not appear in $X_6\cup X_4\setminus\{a\}$ and that $a$
is not a b-vertex.  By (\ref{znob}), we have $b_1\in X_1\cup X_2$.
Let $f(a)=3$.  So $b_3$ is in $N(a)$ and not in $N(b_1)\cup Z$.  Note
that $N(a)=X_3\cup X_5\cup T$.  %
\\
Suppose that $b_1\in X_1$.  So $b_3\in X_3$, and $X_3=\{b_3\}$.  We
may assume that $c(x_6)=2$.  Vertex $b_1$ must have a neighbor $u_2$
of color~$2$, and necessarily $u_2\in X_4$.  Then there is no edge
$xz$ with $x\in X_6$ and $z\in Z$, for otherwise $\{z, x, a_2, b_1,
u_2\}$ induces an $F_1$.  Thus, vertices of $X_6$ are not b-vertices
as they cannot have any neighbor of color~$1$.  It follows that $b_2$
is in $X_1\cup X_4$; in fact it cannot be in $X_1$ by (\ref{eq:2xi})
($X'_1$ cannot contain two b-vertices); so $b_2$ is in $X_4$.  Vertex
$b_1$ must have a neighbor $u_3$ of color~$3$, and necessarily $u_3\in
X_5$.  The definition of $f(a)$ implies that $u_3$ is not a b-vertex,
so it has no neighbor of some color $h\neq 3$.  So color~$h$ does not
appear in $X_1\cup X_4\cup X_6\cup T$, and $h\ge 4$.  Vertex $b_2$
must have a neighbor $u_h$ of color $h$, and (since $|X_3| =1$) we
have $u_h\in X_5$.  But then $\{b_1, a_2, b_3, u_2, a, x_6, u_3,
u_h\}$ induces an $F_{21}$, a contradiction.  \\
Therefore $b_1\in X_2$, so $X_2=\{b_1\}$.  Then (because $b_3$ sees
$a$ and not $b_1$) $b_3\in X_5$, so $X_5=\{b_3\}$.  Vertex $b_1$ must
have a neighbor of color~$3$, so we may assume that $c(a_3)=3$.  The
definition of $f(a)$ implies that $a_3$ is not a b-vertex, so we may
assume that it misses color~$4$.  So color~$4$ does not appear in
$X_6\cup T$, and we may assume $c(x_6)=2$.  Vertex $b_1$ must have a
neighbor $u_4$ of color~$4$, and necessarily $u_4\in X_1\cup X_3$.  If
$u_4\in X_3$, then $|X_3|\ge 2$, so $u_4$ is not a b-vertex by
(\ref{eq:tw235}), and $b_4$ can only be in $X_1$.  Then $b_4$ must
have a neighbor $u_2$ of color~$2$, and $u_2$ can only be in $X_4$.
But then $\{b_1, b_3, b_4, a, u_2, a_3, u_4, x_6\}$ induces an
$F_{21}$, a contradiction.  So color~$4$ does not appear in $X_3$, and
consequently $u_4\in X_1$ and $b_4\in X_1\cup X_4$.  \\
If $b_4$ is in $X_4$, then it has no neighbor of color~$1$, a
contradiction.  So, $b_4$ is in $X_1$.  Vertex $b_4$ must have a
neighbor $u_2$ of color~$2$, and (because of $x_6$) $u_2$ can only be
in $X_4$.  Then there is no edge $xz$ with $x\in X_6$ and $z\in Z$,
for otherwise $\{z, x, b_1, b_4, u_2\}$ induces an $F_1$.  Thus,
vertices of $X_6$ are not b-vertices as they have neither color~$4$
nor a neighbor of color~$4$.
%
%\marginpar{see comment file}
%
It follows that $b_2$ is in $X_1\cup X_4$; in fact it cannot be in
$X_1$ by (\ref{eq:2xi}) ($X'_1$ cannot contain two b-vertices); so
$b_2$ is in $X_4$. Recall that the definition of $f(a)$ implies
that $a$ is the only neighbor of color~$1$ of $a_5=b_3$.  But then
$b_2$ cannot have any neighbor of color~$1$. Thus (\ref{x60})
holds.

%%%%% NOW WE LOOK AT X''_1 AND X''_4

\begin{equation}\label{eq:xs1xs4}
X''_1=\emptyset \mbox{ and } X''_4=\emptyset.
\end{equation}
For suppose on the contrary and up to symmetry that there is a
vertex $x$ in $X''_1$.  Let $c(a_1)=1$.  Every vertex $u\in
X'_1\cup X_3$ satisfies $N(u)\subset N(x)$, so by
Lemma~\ref{lem:compuv}, $X'_1\cup X_3$ contains no b-vertex.  In
particuler $a_1$ is not a b-vertex.  By (\ref{eq:2xi}) and
Lemma~\ref{lem:compuv}, no vertex of $X_1\setminus\{a_1\}$ has
color~$1$.  Thus, and by (\ref{znob}), $b_1$ is in $X_4$, and
since it misses $a_1$ it is in $X'_4$ and has no neighbor in $Z$.
Since $a_1$ is not a b-vertex, color~$f(a_1)$ exists, say
$f(a_1)=2$.  By (\ref{znob}) and the definition of $f(a_1)$, all
b-vertices of color~$2$ must be in $X_2\cup X_4$.  In fact if
$b_2$ is in $X_4$, then by (\ref{eq:2xi}) we have $N(b_1) \subset
N(b_2)$, which contradicts Lemma~\ref{lem:compuv}.  So, $X_4$
contains no b-vertex of color~$2$, and, by (\ref{eq:tw235}), we
have $X_2=\{b_2\}$.  Clearly color~$2$ does not appear in $X_3$,
and by the definition of $f(a_1)$, color~$1$ does not appear in
$X_3$.  Let $c(a_5)=h$.  Then $h$ must appear in $X_3$, for
otherwise $b_2$ cannot have any neighbor of color~$h$.  So we may
assume that $c(a_5)= c(a_3)= 3$.  Therefore $a_3$ and $a_5$ are
the only vertices of color~$3$ in $V\setminus Z$, and since $X_3$
contains no b-vertices
%
%\marginpar{changed slightly}
%
and by (\ref{znob}), we have $b_3=a_5$, and, by (\ref{eq:tw235}),
$|X_5| = 1$.  But now $b_1$ cannot have any neighbor of color~$2$.
Thus (\ref{eq:xs1xs4}) holds.

\begin{equation}\label{eq:ztem}
Z=\emptyset \mbox{ and } T=\emptyset.
\end{equation}
By item~10 of Lemma~\ref{lem:c5} and (\ref{eq:xs1xs4}) there is no
edge between $X_1\cup\cdots\cup X_5$ and $Z$.  So if $Z$ or $T$ is not
empty, then $X_1\cup\cdots\cup X_5$ is a homogeneous set that
contradicts Lemma~\ref{lem:homo}.  Thus (\ref{eq:ztem}) holds.

By the preceding points, each $X_i$ ($i=1, \ldots, 5$) is a stable
set and contains at most one b-vertex, and $ T\cup Z\cup
X_6=\emptyset$. So $b(G)\le 5$.  Moreover, if $a_2, a_3, a_5$ are
b-vertices of three different colors, then, by (\ref{eq:tw235}),
%
%\marginpar{add "by (\ref{eq:tw235})" at 2 places}
%
$a_2$ and $a_3$ cannot have a neighbor of color $c(a_5)$, a
contradiction.  It follows that $b(G)=4$ and that we may assume up
to symmetry that $b_1\in X_1$, $b_4\in X_4$, and $ X_2=\{b_2\}$.
Vertex $b_4$ must have a neighbor $u_2$ of color~$2$, and $u_2$
can only be in $X_5$.  Then, by (\ref{eq:tw235}), $b_3$ must be in
$X_3$, so $X_3=\{b_3\}$. Vertex $b_1$ must have a neighbor $u_3$
of color~$3$, and necessarily, $u_3 \in X_5$.
%
%\marginpar{small change}
%
Vertex $b_2$ must have a neighbor $u_4$ of color~$4$, and $u_4$
can only be in $X_1$. Vertex $b_3$ must have a neighbor $u_1$ of
color~$1$, and $u_1$ can only be in $X_4$.  If both $b_1b_4$ and
$u_1u_4$ are edges, then $\{b_1, b_2, b_4, u_1, u_4\}$ induces an
$F_1$.  If exactly one is an edge, then $\{b_1, b_2, b_3, b_4,
u_1, u_2, u_3, u_4\}$ induces an $F_{19}$.  So both are non-edges.
Then $b_1$ must have a neighbor $v_4$ of color~$4$, which can only
be in $X_4$; and $\{b_1, b_2, b_3, b_4, v_4, u_2, u_3, u_4\}$
induces an $F_{19}$ again.  This completes the proof of
Theorem~\ref{thm:noc5}. $\Box$

%%% SECTION ABOUT BOATS

\section{Graphs that contain a boat}
\label{sec:boats}

Let us call \emph{boat} any graph whose vertex-set can be partitioned
into sets $A_0, A_1, \ldots,$ $A_q,$ $B_0, B_1, \ldots,$ $B_q$ that
satisfy the following properties, where $A=A_0\cup A_1\cup \cdots\cup
A_q$ and $B=B_0\cup B_1\cup \cdots\cup B_q$:
\begin{itemize}
\item
$q\ge 2$ and each of $A_1, \ldots, A_q, B_1, \ldots, B_q$ is not
empty;
\item
If $q=2$ then also $A_0$ and $B_0$ are not empty;
\item
$A_0, A_1, \ldots, A_q$ are pairwise complete to each other,
and $B_0, B_1, \ldots, B_q$ are pairwise complete to each other;
\item
For $j=1, \ldots, q$, $A_j$ is complete to $B_j$ and anticomplete to
$B\setminus B_j$;
\item
$A_0$ is anticomplete to $B_0$.
\end{itemize}
Note that the smallest boats have six vertices: these are the boat
with $q=2$ where each of $A_0, A_1, A_2, B_0, B_1, B_2$ has size one;
and the boat with $q=3$ where each of $A_1, A_2, A_3, B_1, B_2, B_3$
has size one and $A_0=B_0=\emptyset$.  We call these two graphs the
\emph{small boats}.

%%% "BOAT" LEMMA
%
\begin{lemma}\label{lem:boat}
Let $G$ be a graph that contains no $F_1, F_4, F_{10}$ or $C_5$.  If
$G$ contains a boat, then $V(G)$ can be partitioned into sets $M, T,
Z$ such that the subgraph induced by $M$ is a boat and every vertex of
$M$ is complete to $T$ and anticomplete to $Z$.
\end{lemma}
\emph{Proof.} Since $G$ contains a boat, there is a set $M$ of
vertices of $G$ that induces a boat and is maximal with this
property. We use the same notation $A_0, A_1, \ldots, A_q,$ $ B_0,
B_1, \ldots, B_q$ and properties as in the definition of a boat.
Let then $T$ be the set of $M$-complete vertices and $Z$ be the
set of $M$-anticomplete vertices.  In order to prove the lemma, we
need only establish that $V(G)=M\cup T\cup Z$.  Assume the
contrary.  Let $x$ be a vertex of $G$ that is not in $M\cup T\cup
Z$.  Let us fix some notation.  Let $I=\{i\mid 0\le i\le q \mbox{
and } A_i\neq\emptyset\}$.  So $I$ is equal to either $\{0, 1,
\ldots, q\}$ or $\{1, \ldots, q\}$.  Likewise, let $J=\{j\mid 0\le
i\le q \mbox{ and } B_j\neq\emptyset\}$.  Note that we have
$|I\cap J| \ge 3$ (even when $q=2$) by the definition of a boat.
For each $i\in I$, let $a_i$ be an arbitrary vertex in $A_i$ and
$u_i$ be a neighbor of $x$ in $A_i$ (if any).  Likewise, for each
$j\in J$, let $b_j$ be an arbitrary vertex in $B_j$ and $v_j$ be a
neighbor of $x$ in $B_j$ (if any).  We claim that:
\begin{equation}\begin{minipage}{0.8\linewidth}\label{L411}
If there is a pair of integers $i\in I$, $j\in J$, $i\neq j$ such that
$x$ has a neighbor in each of $A_i, B_j$, then $x$ is complete to
either $A\setminus(A_i\cup A_j)$ or $B\setminus(B_i\cup B_j)$.
\end{minipage}
\end{equation}
Note that vertices $u_i$ and $v_j$ exist.  If $x$ has non-neighbors
$u'\in A\setminus(A_i\cup A_j)$ and $v'\in B\setminus(B_i\cup B_j)$,
then $\{u', u_i, x, v_j, v'\}$ induces an $F_1$ or a $C_5$, a
contradiction.  Thus (\ref{L411}) holds.

\begin{equation}\begin{minipage}{0.8\linewidth}\label{L412}
There is no pair of integers $i, j\in I\cap J$, $i\neq j$ such that
$x$ has a neighbor in each of $A_i, A_j, B_i, B_j$.
\end{minipage}
\end{equation}
For suppose the contrary; so vertices $u_i, u_j, v_i, v_j$ exist.
Since $i,j$ play symmetric roles, we may assume that $i\neq 0$, so
$u_i$ and $v_i$ are adjacent.  By (\ref{L411}) and up to symmetry, we
may assume that $x$ is complete to $A\setminus(A_i\cup A_j)$.
Consider any $h\in (I\cap J)\setminus\{i, j\}$, which is not empty.
We know that $x$ is complete to $A_h$, and $B_h\neq \emptyset$.  Then
$x$ is adjacent to $b_h$, for otherwise $\{a_h, u_i, x, v_i, v_j,
b_h\}$ induces an $F_4$ (if $h=0$) or $F_{10}$ (if $h\neq 0$).  So $x$
is complete to $B_h$.  We can repeat this argument for every pair of
integers from $I\cap J$, and it follows that $x$ is complete to
$A_\ell\cup B_\ell$ for every $\ell\in I\cap J$, in particular for
$1\le\ell\le q$.  Thus, if $0\in I\cap J$, we obtain $x\in T$, a
contradiction.  So $0\notin I\cap J$, say $B_0=\emptyset$, and so
$q\ge 3$.  Since $x$ is not in $T$, it has a non-neighbor $w$ in
$A\cup B$, and it must be that $w\in A_0$.  But then $\{w, a_1, a_2,
x, b_1, b_3\}$ induces an $F_4$.  Thus (\ref{L412}) holds.

\begin{equation}\begin{minipage}{0.8\linewidth}\label{L413}
There do not exist two pairs of integers $g, h\in I$, $g\neq h$ and
$i, j\in J$, $i\neq j$ such that $x$ has a neighbor in each of $A_g,
A_h, B_i, B_j$.
\end{minipage}
\end{equation}
For suppose the contrary; so vertices $u_g, u_h, v_i, v_j$ exist.
First suppose that $\{g, h\}\cap \{i, j\}\neq\emptyset$, say
$g=i$. By (\ref{L412}) we may assume that $h\neq j$.  If $g=i=0$,
then $A_j$ and $B_h$ are not empty, and by (\ref{L412}), $x$ has
no neighbor in those two sets; but then $\{a_j, u_0, x, v_0,
b_h\}$ induces an $F_1$. So $g=i\neq 0$, and $u_g$ sees $v_i$. One
of $h, j$ is not equal to $0$, say $j\neq 0$.  So
$A_j\neq\emptyset$ and, by (\ref{L412}), $x$ has no neighbor in
that set; but then $\{a_j, u_g, u_h, x, v_i, v_j\}$ induces an
$F_{10}$.  Now we may assume that the four integers $g, h, i, j$
are different.  We may assume that none of $h, i, j$ is equal to
$0$; so $A_j$ and $B_h$ are not empty. If $x$ has no neighbor in
those two sets, then $\{a_j, u_g, x, v_i, b_h\}$ induces an $F_1$.
Now, by (\ref{L412}), $x$ has a neighbor in $A_j$, or in $B_h$,
but not in both.
%
%\marginpar{{\it see comment file}}
%
If $x$ has a neighbor in $B_h$, then $\{a_j, u_h, u_g, x , v_h,
v_j\}$ induces an $F_{10}$. If $x$ has a neighbor in $A_j$, then
$\{b_h, v_j, v_i, x , u_h, u_j\}$ induces an $F_{10}$. Thus,
(\ref{L413}) holds.

\begin{equation}\begin{minipage}{0.8\linewidth}\label{L414}
There is no pair of integers $i\in I$, $j\in J$, $i\neq j$ such that
$x$ has a neighbor in each of $A_i, B_j$.
\end{minipage}
\end{equation}
For suppose the contrary; so vertices $u_i$ and $v_j$ exist.  By
(\ref{L411}) and up to symmetry, we may assume that $x$ is
complete to $A\setminus(A_i\cup A_j)$.  Thus $x$ has neighbors in
at least two of the sets $A_0, \ldots, A_q$, and so, by
(\ref{L413}), it has no neighbor in $B\setminus B_j$.  Consider
any $h\in I\cap J\setminus \{i,j\}$.  So $A_h$ and $B_h$ are not
empty, and $x$ is complete to $A_h$ and anticomplete to $B_h$.  We
claim that there is a non-neighbor $w$ of $x$ in
$B\setminus(B_j\cup B_h)$; indeed, if $B_i\neq\emptyset$, then we
can take any $w\in B_i$; and if $B_i=\emptyset$, then $i=0$, $q\ge
3$, so there is an integer $g\in (I\cap J)\setminus \{h, j\}$, and
we can take any $w\in B_g$.  Now we can apply (\ref{L411}) to $h$
and $j$, and the existence of $w$ implies that $x$ is complete to
$A\setminus (A_j\cup A_h)$.  In summary, we have established that
$x$ is complete to $A\setminus A_j$ and anti-complete to
$B\setminus B_j$.  If $x$ is complete to $B_j$, then we can add
$x$ to $A_j$ and obtain a boat (with sets $A_0, \ldots,
A_j\cup\{x\}, \ldots, A_q, B_0, \ldots, B_q$) that contradicts the
maximality of $M$.  Therefore $x$ has a non-neighbor $w_j$ in
$B_j$.  Suppose that $j\neq 0$; so $A_j\neq\emptyset$ and $a_j$
sees both $v_j, w_j$.  Then $w_j$ sees $v_j$, for otherwise
$\{w_j, b_h, v_j, x, a_i\}$ induces an $F_1$; and $x$ sees $a_j$,
for otherwise $\{b_h, w_j, a_j, a_i, x\}$ induces an $F_1$; but
then $\{a_h, a_j, x, v_j, w_j, b_h\}$ induces an $F_4$ (if $h=0$)
or $F_{10}$ (if $h\neq 0$), a contradiction.  So $j=0$.  If $x$
has a non-neighbor $y\in A_0$, then $\{y, a_i, x, v_0, b_h\}$
induces an $F_1$.  Thus $x$ is complete to $A_0$.  Set
$A_{q+1}=\{x\}$, $B_{q+1}=B_0\cap N(x)$ and $B'_0 = B_0\setminus
N(x)$.  Note that $v_j\in B_{q+1}$ and $w_j\in B'_0$.  Moreover,
every $v\in B_{q+1}$ sees every $w\in B'_0$, for otherwise $\{a_i,
x, v, b_h, w\}$ induces an $F_1$.  But now we find a larger boat,
with sets $A_0, \ldots, A_q, A_{q+1}, B'_0, B_1, \ldots, B_q,
B_{q+1}$,
%
%\marginpar{{\it see comment file}}
%
which contradicts the maximality of $M$. Thus (\ref{L414}) holds.

\medskip

Since $x\notin Z$, up to symmetry we may assume that $x$ has a
neighbor $u_h\in A_h$ for some $h$ with $0\le h\le q$.  By
(\ref{L414}), $x$ has no neighbor in $B\setminus B_h$.  Consider
any $i\in (I\cap J)\setminus \{h\}, i \not= 0,$
%
%\marginpar{see comment file}
%
and suppose that $x$ has a non-neighbor $w_i$ in $A_i$.  Pick any
$j\in J\setminus\{h, i\}$. Then $\{x, u_h, w_i, b_i, b_j\}$
induces an $F_1$, a contradiction. Thus $x$ is complete to $A_i$.
By repeating this argument with $i$ instead of $h$, we obtain that
$x$ is complete to $A\setminus A_0$, and by (\ref{L414}) it is
anti-complete to $B$.  But now we find a larger boat, with sets
$A_0\cup\{x\}, A_1, \ldots, A_q, B_0, B_1, \ldots, B_q$, which
contradicts the maximality of $M$.  This completes the proof of
Lemma~\ref{lem:boat}.  $\Box$

%%%%% THEOREM about BOATS

\begin{theorem}\label{thm:noboat}
Let $G$ be a minimal counterexample to Conjecture~\ref{cj1}.  Then $G$
contains no boat.
\end{theorem}
\emph{Proof.} By Theorem~\ref{thm:noc5}, $G$ contains no $C_5$.
Suppose that $G$ contains a boat.  Then, by Lemma~\ref{lem:boat},
the vertex-set of $G$ can be partitioned into sets $M, T, Z$ such
that $M$ induces a boat and every vertex of $M$ is $T$-complete
and $Z$-anticomplete.  Then $T$ and $Z$ are empty, for otherwise
$M$ is a homogeneous set that contradicts Lemma~\ref{lem:homo}.
Thus $V(G)=M$. We use the same notation $A_0, A_1, \ldots,$ $
A_q,$ $ B_0, B_1, \ldots, B_q$ and properties as in the definition
of a boat.  As in Lemma~\ref{lem:boat}, let $I=\{i\mid 0\le i\le q
\mbox{ and } A_i\neq\emptyset\}$, $J=\{j\mid 0\le i\le q \mbox{
and } B_j\neq\emptyset\}$, and note that $|I\cap J| \ge 3$.  For
each $i\in I$, let $a_i$ be an arbitrary vertex in $A_i$, and for
each $j\in J$, let $b_j$ be an arbitrary vertex in $B_j$.

Let $c$ be a b-coloring of $G$ with $k=b(G)>\chi(G)$ colors.  For each
color $\ell\in\{1, \ldots, k\}$ let $d_\ell$ be a b-vertex of color
$\ell$, and let $D=\{d_1, \ldots, d_k\}$.

By the definition of a boat each of $A_0, A_1, \ldots, A_q,$ $
B_0, B_1, \ldots, B_q$ is a homogeneous set, so it satisfies the
properties described in Lemma~\ref{lem:homo}.  It follows (recall
Lemma~\ref{lem:compuv}) that:
\begin{equation}\begin{minipage}{0.8\linewidth}\label{t4h}
Each $A_i$ ($i\in I$) is a clique or a stable set.  All vertices of
$A$ have different colors.  Any two non-adjacent vertices of $A$ are
not b-vertices.  The same holds for $B$.
\end{minipage}
\end{equation}
We will now prove
\begin{equation}\begin{minipage}{0.8\linewidth}\label{t4d}
If two of the sets $A_1, \ldots, A_q$ contain a member of $D$, then
$A$ contains vertices of all colors.  The same holds for $B$.
\end{minipage}
\end{equation}
For suppose up to symmetry that $d_1\in A_1$ and $d_2\in A_2$.
Consider any color $\ell$ that appears in $B$.  If $\ell$ appears in
$B\setminus B_1$, then it does not appear in $B_1$, and since $d_1$
must have a neighbor of color $\ell$, such a neighbor must be in $A$.
If $\ell$ appears in $B_1$, then it does not appear in $B_2$, and
since $d_2$ must have a neighbor of color $\ell$, such a neighbor must
be in $A$.  So we have established that all colors that appear in $B$
also appear in $A$; and so, all colors appear in $A$.  Thus
(\ref{t4d}) holds.

\begin{equation}\begin{minipage}{0.8\linewidth}\label{t4a1}
If $A_i$ is not a clique for some $i\in \{1, \ldots, q\}$, then
every $B_j$ with $j\in \{1, \ldots, q\}\setminus \{i\}$ is a
stable set. The same holds with $A$ and $B$ interchanged.
\end{minipage}
\end{equation}
For suppose on the contrary, and up to symmetry, that there are two
non-adjacent vertices $u,v$ in $A_1$ and two adjacent vertices $x,y$
in $B_2$.  Let $h=3$ if $q\ge 3$ and $h=0$ if $q=2$.  Then $\{a_h, u,
v, a_2, b_1, x, y\}$ induces an $F_{11}$, a contradiction.  Thus
(\ref{t4a1}) holds.

\begin{equation}\label{t4aibi}
\mbox{For each $i\in\{0, \ldots, q\}$, one of $A_i, B_i$ is a clique.}
\end{equation}
For suppose on the contrary that there exist non-adjacent vertices
$u,v\in A_i$ and non-adjacent vertices $x,y\in B_i$.  Pick two
integers $h, j$ in $(I\cap J)\setminus \{i\}$.  Then $\{u, v, x, y,
a_h, a_j, b_h, b_j\}$ induces an $F_{12}$ (if $i=0$) of $F_{14}$ (if
$h=0$ or $j=0$) or $F_{15}$ (if $h, i, j>0$).  Thus (\ref{t4aibi})
holds.

\begin{equation}\label{t4k1}
\mbox{One of $A, B$ is a clique.}
\end{equation}
For suppose the contrary.  So one of the $A_i$'s ($i\in I$) is not a
clique and one of the $B_j$'s ($j\in J$) is not a clique.  By
(\ref{t4aibi}) and up to symmetry, we may assume that $A_1$ is not a
clique and one of $B_0, B_2$ is not a clique.  By (\ref{t4a1}) and
(\ref{t4aibi}), $B_1$ is a clique and each of $B_2, \ldots, B_q$ is a
stable set.  Note that this implies, by (\ref{t4h}), that each of
$B_2, \ldots, B_q$ contains at most one b-vertex (and if it contains
one, then it has size one).  Let $u,v$ be two non-adjacent vertices in
$A_1$, and let $x,y$ be two non-adjacent vertices in $B_0$ or $B_2$.
By (\ref{t4h}), $u$ and $v$ are not b-vertices, and we may assume that
they have color $1$ and $2$.  By (\ref{t4h}) again, $d_1, d_2$ are in
$B$, each of them is adjacent to all other vertices of $B$, and
clearly they are not in $B_1$.  Suppose that $B_0$ is not a clique,
say $x,y$ are in $B_0$.  If $q\ge 3$, then $\{u, v, a_2, a_3, b_2,
b_3, x, y\}$ induces an $F_{12}$.  So $q=2$, and so $A_0\neq
\emptyset$.  Note that $d_1, d_2$ are not both in $B_2$ (because $B_2$
can contain at most one b-vertex).  So we may assume that $d_1\in
B_0$.  But $d_1$ is adjacent to $x$, a contradiction to (\ref{t4h}).
%
%\marginpar{see comment file}
%
%But then $\{a_0, u, v, a_2, b_2, d_1, x, y\}$
%induces an $F_{7}$.
%
Therefore $B_0$ is a clique.  So $x,y$ are in $B_2$, which restores
the symmetry between $A$ and $B$.  Thus $A_2$ is a clique, each of
$A_1, A_3, \ldots, A_q$ is a stable set and contains at most one
b-vertex, and $A_0$ is a clique.  If $q\ge 4$, then $\{u, v, a_3, a_4,
b_3, b_4, x, y\}$ induces an $F_{12}$.  So $q\le 3$.  None of $d_1,
d_2$ is in $B_2$ (because $B_2$ is now a stable set of size at least
two, so it does not contain any b-vertex), so they are in $B_0\cup
B_3$.  Since $B_3$ is a stable set, it contains at most one of $d_1,
d_2$, and so at least one of these is in $B_0$, say $d_2\in B_0$.  By
symmetry, we may assume that $x,y$ have color $3$ and $4$
respectively, vertices $d_3, d_4$ are in $A_0\cup A_3$, vertex $d_4$
is in $A_0$, and each of $d_3, d_4$ is adjacent to all other vertices
of $A$.  But then $\{u, v, d_3, d_4, x, y, d_1, d_2\}$ induces an
$F_{7}$ (if $d_3\in A_3$ and $d_1\in B_3$) or an $F_{6}$ (else).  Thus
(\ref{t4k1}) holds.

\begin{equation}\label{t4k2}
\mbox{Both $A, B$ are cliques.}
\end{equation}
For suppose the contrary.  By (\ref{t4k1}), we may assume that $A$ is
not a clique and $B$ is a clique.  The vertices of $B_0$ are
simplicial, so they are not b-vertices by Lemma~\ref{lem:simp}.
Moreover, if two of the sets $B_1, \ldots, B_q$ contain a b-vertex,
then, by (\ref{t4d}), $B$ contains vertices of all colors, and so $G$
has a clique of size $k$, a contradiction.  Therefore we may assume up
to symmetry that $B\cap D\subseteq B_1$.  Let $u,v$ be two
non-adjacent vertices in $A$.  By (\ref{t4h}), $u,v$ are not
b-vertices and we may assume that they have color $1$ and $2$.  Then
vertices $d_1, d_2$ are not in $A$, so they are in $B_1$.
%
%\marginpar{see comment file}
%
If $u,v \in A_i, i \not=0, i\not=1$, then by (24), $B_1$ is a
stable set of size at least two, a contradiction to the assumption
that $B$ is a clique. So, $u,v$ are not in $A\setminus A_0$, and
this argument shows that $A\setminus A_0$ is a clique.  Thus $u,v$
are in $A_0$. Now $A_1\cup B_1$ is a clique, so we may assume that
it does not contain any vertex of color $3$. Since $d_1, d_2$ are
b-vertices, they must have a neighbor $x_3$ of color $3$, which
must be in $B\setminus B_1$. Vertex $x_3$ is not a b-vertex
(because $B\cap D\subseteq B_1$), so $d_3$ is in $A$, and
$d_3\notin A_1$ by the choice of color $3$.  By (\ref{t4h}), $d_3$
is adjacent to all of $A\setminus\{d_3\}$.  But then $\{u, v, d_3,
a_1, d_1, d_2, x_3\}$ induces an $F_5$.  Thus (\ref{t4k2}) holds.

\medskip

Since $A$ is a clique, there is a color, say color $1$, that does not
appear in $A$.  So $d_1$ is in $B$ and is the unique vertex of color
$1$.  Likewise, there is a color, say color $2$, that does not appear
in $B$, and so $d_2$ is in $A$ and is the unique vertex of color $2$.
Since $d_1$ must have a neighbor of color $2$, vertices $d_1, d_2$ are
adjacent, and so we may assume that $d_1\in B_1$ and $d_2\in A_1$.  By
(\ref{t4d}) we have $A\cap D\subseteq A_1$ and $B\cap D\subseteq B_1$.
But then $A_1\cup B_1$ is a clique that contains vertices of all
colors, a contradiction.  This concludes the proof of the theorem.
$\Box$

%%%%% END SECTION "BOAT"

%%%% FINAL SECTION ???

\section{Proof of the main result}

Suppose that Theorem~\ref{thm:main} fails.  So there is a minimal
counterexample $G$ to Conjecture~\ref{cj1}.  By
Theorems~\ref{thm:noc5} and~\ref{thm:noboat}, $G$ contains no
$C_5$ and no boat.  Since $G$ contains no $C_5$, no $F_1$ ($=P_5$)
and no $F_{10}$ ($=\overline{P}_6$), it contains no odd hole and
no odd antihole, so it is \emph{perfect} \cite{CRST}, that is, $G$
satisfies $\chi(H)=\omega(H)$ for every induced subgraph $H$ of
$G$. (Actually, since $G$ also contains no $\overline{C}_6$ (which
is a boat), it is \emph{weakly chordal} (i.e., it contains no hole
and no antihole of length at least five), and reference \cite{hay}
implies the perfectness of $G$ more simply than \cite{CRST}.)

Let $c$ be a b-coloring of $G$ with $k=b(G)>\chi(G)$ colors.  For each
color $i\in\{1, \ldots, k\}$, let $d_i$ be a b-vertex of color $i$,
and let $D=\{d_1, \ldots, d_k\}$.  Note that $G$ contains no clique of
size $k$, for otherwise we would have $k\le \omega(G)\le \chi(G) <
b(G) = k$, which is impossible.  In particular $D$ is not a clique.

We observe that:
\begin{equation}\label{e2k2}
\mbox{$G$ contains a $2K_2$.}
\end{equation}
For suppose that $G$ contains no $2K_2$.  Since $D$ is not a clique,
we may assume without loss of generality that $d_1, d_2$ are not
adjacent.  Since $d_1$ is a b-vertex, it has a neighbor $x_2$ of color
$2$.  Since $d_2$ is a b-vertex, it has a neighbor $x_1$ of color $1$.
Then $x_1, x_2$ are adjacent, for otherwise $\{d_1, d_2, x_1, x_2\}$
induces a $2K_2$.  Since $d_1$ is a b-vertex, by
Lemma~\ref{lem:compuv} we cannot have $N(d_1)\subseteq N(x_1)$; so
there exists a vertex $u$ that is adjacent to $d_1$ and not to $x_1$.
Likewise, there exists a vertex $v$ that is adjacent to $d_2$ and not
to $x_2$.  Then $u$ is adjacent to $d_2$, for otherwise $\{u, d_1,
d_2, x_1\}$ induces a $2K_2$; and $u$ is adjacent to $x_2$, for
otherwise $\{u, d_1, d_2, x_1, x_2\}$ induces a $C_5$.  Likewise, $v$
is adjacent to $d_1$ and $x_1$.  But now $\{u, v, d_1, d_2, x_1,
x_2\}$ induces a boat (if $u, v$ are not adjacent) or an $F_{10}$ (if
$u, v$ are adjacent), a contradiction.  Thus (\ref{e2k2}) holds.

\medskip

Since $G$ contains a $2K_2$, there is a subset $B$ of $V(G)$ such that
the subgraph induced by $B$ has at least two components, each
component of $B$ has at least two vertices, and $B$ is maximal with
this property.  Let $B_1, \ldots, B_r$ be the components of $B$, with
$r\ge 2$.  Let $S$ be the set of vertices of $V\setminus B$ that are
$B$-anticomplete.  Note that $S$ is a stable set, for otherwise two
adjacent vertices of $S$ could be added to $B$, which would contradict
the maximality of $B$.  Let $A=V\setminus (B\cup S)$.  We claim that:
\begin{equation}\begin{minipage}{0.85\linewidth}\label{b1}
$r=2$ and there is a component of $B$, say $B_2$, such that $B_2$ is a
clique and every vertex of $A$ is $B_2$-complete and has a neighbor in
$B_1$.
\end{minipage}
\end{equation}
Consider any vertex $a\in A$.  By the definition of $S$, $a$ has a
neighbor in $B$.  If $a$ has no neighbor in some component $B_j$ of
$B$, then $B\cup\{a\}$ contradicts the maximality of $B$ (as every
component of $B\cup \{a\}$ has size at least two and $B$ has at least
two components, one that contains $a$ and one that includes $B_j$).
So $a$ has a neighbor in each component of $B$.  If every vertex of
$A$ is $B$-complete, then $B$ is a homogeneous set, which contradicts
Lemma~\ref{lem:homo}.  So there is a vertex $a_0$ of $A$ that is not
$B$-complete, say $a_0$ has a non-neighbor in $B_1$.  Since $B_1$ is
connected, there are adjacent vertices $u_1, v_1$ in $B_1$ such that
$a_0$ is adjacent to $u_1$ and not to $v_1$.  If $a_0$ also has a
non-neighbor in another component $B_i$ of $B$ ($i\neq 1$), then there
are adjacent vertices $u_i, v_i$ in $B_i$ such that $a_0$ is adjacent
to $u_i$ and not to $v_i$, and $\{a_0, u_1, v_1, u_i, v_i\}$ induces
an $F_1$.  Therefore only component $B_1$ of $B$ contains a
non-neighbor of $a_0$, that is, $a_0$ is $B\setminus B_1$-complete.
Consider any other vertex $a'_0$ in $A$ that is not $B$-complete.
Just like for $a_0$, there is a component $B_i$ and adjacent vertices
$u_i, v_i$ in $B_i$ such that $a'_0$ is adjacent to $u_i$ and not to
$v_i$ and $a'_0$ is $B\setminus B_i$-complete.  If $i\neq 1$, then
$\{a_0, a'_0, u_1, v_1, u_i, v_i\}$ induces an $F_4$ or a boat, a
contradiction.  So $i=1$ for each $a'_0$.  This implies that every
vertex of $A$ is $B\setminus B_1$-complete, so $B\setminus B_1$ is
homogeneous, and Lemma~\ref{lem:homo} implies that $r=2$ and $B_2$ is
a clique.  Thus (\ref{b1}) holds.

\begin{equation}\begin{minipage}{0.85\linewidth}
\label{eq:aut1} For each $a\in A$ and each component $C$ of
$B_1\setminus N(a)$, every vertex in $B_1\setminus C$ is either
$C$-complete or $C$-anticomplete.  Furthermore, there is a vertex in
$B_1 \setminus C$ that is ($C\cup\{a\}$)-complete.
\end{minipage}
\end{equation}
Suppose that some vertex $z$ in $B_1 \setminus C$ has a neighbour and
a non-neighbour in $C$.  Then $z$ is adjacent to $a$ by the definition
of $C$.  Since $C$ is connected, there are adjacent vertices $y,y' \in
C$ such that $z$ sees $y$ and misses $y'$.  Let $t$ be any vertex in
$B_2$.  Then $\{t, a, z, y, y'\}$ induces an $F_1$, a contradiction.
Furthermore, since $B_1$ is connected, some vertex in $B_1 \cap N(a)$
must have a neighbour in $C$, and so is ($C\cup\{a\}$)-complete.  Thus
(\ref{eq:aut1}) holds.

\begin{equation}\begin{minipage}{0.85\linewidth}\label{eq:aut2}
For every $a\in A$, each component of $B_1\setminus N(a)$ is a
homogeneous set and a clique.
\end{minipage}
\end{equation}
Let $C$ be any component of $B_1\setminus N(a)$.  Suppose that $C$ is
not homogeneous.  So there are adjacent vertices $x,y$ in $C$ and a
vertex $u$ not in $C$ that sees $x$ and misses $y$.  By
(\ref{eq:aut1}), we have $u \not\in B_1$ and so $u \in A$.  Let $t$ be
any vertex in $B_2$.  Then $u$ sees $a$, for otherwise $\{a, t, u, x,
y\}$ induces an $F_1$.  By (\ref{eq:aut1}), there is a vertex $z$ in
$B_1 \cap N(a)$ that is ($C\cup\{a\}$)-complete.  But now $\{a, t, u,
x, y, z\}$ induces a boat or an $F_4$, a contradiction.  So $C$ is
homogeneous.  Then Lemma~\ref{lem:homo} implies that $C$ is a clique.
Thus (\ref{eq:aut2}) holds.

\begin{equation}\label{di0}
\mbox{$S$ contains no b-vertex.}
\end{equation}
Indeed, if $x$ is any vertex in $S$ and $t$ is any vertex in $B_2$,
then $N(x)\subset N(t)$ and Lemma~\ref{lem:compuv} implies that $x$
cannot be a b-vertex.  Thus (\ref{di0}) holds.

Let $G'$ be the graph obtained from $G$ by removing all edges whose
two endvertices are in $B_2$.
\begin{equation}\label{eq:gpf}
\mbox{$G'$ does not contain any $C_5$, boat, or member of $\cal F$.}
\end{equation}
For suppose that $G'$ has an induced subgraph $F$ that is either a
$C_5$, a boat, or a member of $\cal F$.  If $F$ is a boat, we may
assume that it is a small boat, since every boat contains a small boat.
If $F$ contains at most one vertex of $B_2$, then $F$ is an induced
subgraph of $G$, a contradiction.  So $F$ must contain at least two
vertices of $B_2$.  Then these vertices are pairwise non-adjacent
twins in $F$, which implies that $F$ is not a $C_5$ or a small boat
(since such graphs do not have any pair of non-adjacent twins); more
precisely $F$ is one of $F_2, F_3, F_5, F_6, F_7, F_9, F_{11}, F_{12},
F_{13}, F_{14}, F_{15}, F_{19}, F_{21}$ and it has exactly two
vertices of $B_2$.  In fact $F$ cannot be any of $F_{19}, F_{21}$, for
that would imply that $G'$ contains a $C_5$, which we have already
excluded.  Likewise, $F$ cannot be any of $F_{12}, F_{14}, F_{15}$,
since that would imply that $G'$ contains a boat, which is also
excluded.  Therefore $F$ is one $F_2, F_3, F_5, F_6, F_7, F_9, F_{11},
F_{13}$. \\
Suppose that $F$ is either $F_2$ or $F_3$.  So $F$ has vertices $x, y,
a, z_1, \ldots, z_p$, with $x,y\in B_2$, and edges $xa, ya$, and
either (if $F$ is $F_2$) $p=4$ and $\{z_1, \ldots, z_4\}$ induces a
$P_4$, or (if $F$ is $F_3$) $p=6$ and $\{z_1, \ldots, z_6\}$ induces a
$2P_3$.  Since $x,y$ are in $B_2$, vertices $z_1, \ldots, z_p$ must be
in $B_1$ and $a$ must be in $A$.  But then $a, z_1, \ldots, z_p$
contradict Claim (\ref{eq:aut2}). \\
Suppose that $F$ is either $F_5$ or $F_9$.  So $F$ has vertices $x, y,
a, b, z_1, \ldots, z_p$, with $x,y\in B_2$, and edges $xa, xb,$ $ya,
yb,$ $ab,$ $az_1,$ $z_1z_2,$ $z_1z_3,$ $z_2z_3$ and either (if $F$ is
$F_5$) $p=3$ and $az_2$ is an edge, or (if $F$ is $F_9$) $p=6$ and
vertices $z_4, z_5, z_6$ induce a $P_3$ and are adjacent to $a$.
Since $x,y$ are in $B_2$, vertices $z_1, \ldots, z_p$ must be in $B_1$
and $a, b$ must be in $A$.  But then $b, z_1, z_2, z_3$ contradict
Claim (\ref{eq:aut2}). \\
Suppose that $F$ is either $F_6$ or $F_7$.  So $F$ has vertices $x,
y,$ $a, b,$ $z_1, \ldots,$ $z_4$, with $x,y\in B_2$, and edges $xa,
xb, ya, yb, ab, z_1z_2, z_1z_3, z_1z_4, z_2z_3, z_2z_4$ and (if $F$ is
$F_7$) the edge $az_1$.  Since $x,y$ are in $B_2$, vertices $z_1,
\ldots, z_4$ must be in $B_1$ and $a, b$ must be in $A$.  But then $b,
z_2, z_3, z_4$
%
%\marginpar{need the z to be non-clique}
%
contradict Claim (\ref{eq:aut2}). \\
Suppose that $F$ is $F_{11}$.  So $F$ has vertices $x, y, a, b,
z_1, z_2, w$, with $x,y\in B_2$, and edges $xa, xb,$ $ya, yb,$
$ab,$ $az_1, az_2,$ $z_1z_2,$ $z_1w,$ $z_2w$, $xw, yw$.  Since
$x,y$ are in $B_2$, vertices $z_1, z_2$ must be in $B_1$ and $a,
b, w$ must be in $A$.  By Claim (\ref{eq:aut1}), there is a vertex
$z$ in $B_1$ that is adjacent to $b, z_1, z_2$.  Then $z$ sees
$w$, for otherwise $\{z, b, x, w, z_1\}$ induces a $C_5$.  But
then $\{a, b, x, w, z, z_1\}$ induces a
small boat ($\overline{C}_6$) or $F_{10}$ in $G$, a contradiction.  \\
Finally suppose that $F$ is an $F_{13}$.  So $F$ has vertices $x, y,
a, b, u, v, z_1, z_2$, with $x,y\in B_2$, and edges $ab, ax, ay, bx,
by, xu, xv, yu, yv, uz_1, uz_2, vz_1, vz_2, z_1z_2$.  Since $x,y$ are
in $B_2$, vertices $z_1, z_2$ must be in $B_1$ and $a, b, u, v$ must
be in $A$.  By Claim (\ref{eq:aut1}), there is a vertex $z$ in $B_1$
that is adjacent to $b, z_1, z_2$.  Vertex $z$ sees $u$, for otherwise
$\{z, b, x, u, z_1\}$ induces a $C_5$.  Similarly $z$ sees $v$.  Then
$z$ sees $a$, for otherwise $\{a, b, x, u, z, z_1\}$ induces a boat.
But then $\{a, b, x, z, u, v, z_1\}$ induces an $F_{11}$ in $G$, a
contradiction.  Thus (\ref{eq:gpf}) holds.

\begin{equation}\begin{minipage}{0.85\linewidth}\label{eq:b2b}
Some vertex $d$ in $B_2$ is the unique b-vertex of $G$ of color
$c(d)$.
\end{minipage}
\end{equation}
%\marginpar{\it{changed}}
%
Suppose for each b-vertex $d$ in $B_2$, there is a b-vertex $d'$
of the same color. Then $d' \not\in B_2$ ($B_2$ is a clique); so,
$c$ is a b-coloring of $G'$ with $k$ colors, and $G'$ is a smaller
counterexample than $G$, a contradiction. Thus, (\ref{eq:b2b})
holds.

Since the vertices of $B_2$ are pairwise adjacent twins,
(\ref{eq:b2b}) implies that they are all b-vertices, and we may
assume that $B_2= \{d_1, \ldots, d_{\ell}\}$, with $\ell \ge 2$.

Next, we will prove
\begin{equation}\begin{minipage}{0.75\linewidth}\label{eq:b1}
$B_1$ contains a vertex $x_i$ of color $i$, $ 1\leq i\leq \ell$, that
is not a b-vertex of $G$.
\end{minipage}
\end{equation}

Let $S_1, \ldots, S_k$ be the color classes of the b-coloring $c$,
with $d_i\in S_i$ for each $i=1, \ldots, \ell$.  We have $k = b(G) >
\chi(G) = \omega(G)$ since $G$ is perfect.  So $k-1\ge \omega(G)$.
Note that $S_2, \ldots, S_k$ form a b-coloring of $G\setminus S_1$; so
$b(G\setminus S_1) \ge k - 1$.  Since $G$ is minimally b-imperfect, we
have $b(G\setminus S_1) = \chi(G\setminus S_1) = \omega(G\setminus
S_1)$.  Combining the above inequalities, we get $k - 1 \ge \omega(G)
\ge \omega(G\setminus S_1) = \chi(G\setminus S_1) = b(G\setminus S_1)
\ge k-1$.  So, equality must hold throughout, in particular we have
$\omega(G\setminus S_1) =\omega(G) = k - 1$.  So $G\setminus d_1$
contains a clique $K$ of size $\omega(G)=k-1$.  If $K$ contains a
vertex $x$ of $S$, then $(K\setminus x)\cup \{d_1, d_2\}$ is a clique
of size $k$ in $G$, a contradiction.  If $K$ contains no vertex of
$B_1$, then we have $K\subseteq A\cup B_2\setminus d_1$, and then
$K\cup \{d_1\}$ is a clique of size $k$ in $G$, again a contradiction.
So $K$ contains a vertex of $B_1$ and $K\subseteq B_1\cup A$.  Then we
have $|K\cap A|\le k-1-\ell$, for otherwise $(K\cap A)\cup B_2$ would
be a clique of size at least $k$.  Consequently, $K\cap B_1$ has size
at least $\ell$ and at least $\ell-1$ of the colors $1, \ldots, \ell$
appear in $K\cap B_1$.  So we may assume up to symmetry that $B_1$
contains a vertex $x_i$ of color $i$, $1 \leq i \leq \ell$.  If $x_i$
is not a b-vertex, then we are done.  Suppose $x_i$ is a b-vertex.
So, it must have neighbors $x_j$ of color $j$, for all $j \in \{1,
\ldots, \ell\} \setminus \{i\}$.  The vertices $x_j$ are in $B_1$
necessarily.  By (\ref{eq:b2b}), some such $x_j$ is not a b-vertex.
So, (\ref{eq:b1}) holds.

For simplicity, let $x_1$ be the vertex of color $1$ in $B_1$ that is
not a b-vertex.  There must be a color $m$ such that $x_1$ is the only
neighbor of color $1$ of every b-vertex of color $m$.  Thus $d_m$ is
not adjacent to $d_1$, so it is not in $A$; therefore, it is in $B_1$.
Moreover $m>\ell$.  Vertices $d_1, \ldots, d_\ell$ must have a
neighbor $u_m$ of color $m$, and clearly $u_m$ is in $A$.  Moreover
$u_m$ is not a b-vertex (because it is adjacent to $d_1$ and by the
property of $x_1$), so there is a color $n\neq 1$ such that $u_m$ has
no neighbor of color $n$.  Thus $n>\ell$.  Vertices $d_1, \ldots,
d_{\ell}$ must have a neighbor $u_n$ of color $n$, and clearly $u_n$
is in $A$.  Let $C$ be the component of $B_1\setminus N(u_m)$ that
contains $d_m$.  By (\ref{eq:aut2}), $C$ is a homogeneous set and a
clique.

\begin{equation}\label{eq:dmk}
\mbox{$N(d_m)\cap B_1$ is a clique.}
\end{equation}
Suppose on the contrary that $d_m$ has two neighbors $x, y$ in
$B_1$ that are not adjacent.  Since $C$ is a homogeneous set and a
clique, vertices $x, y$ are in $B_1\setminus C$ and so they are
both adjacent to $u_m$. Then $u_n$ sees $x$, for otherwise $\{u_n,
d_1, u_m, x, d_m\}$ induces an $F_1$ or $C_5$.  Likewise $u_n$
sees $y$.  Then $u_n$ misses $d_m$, for otherwise $\{d_1, d_2,
u_n, u_m, d_m, x, y\}$ induces an $F_{11}$. Since this holds for
every vertex $u_n$ of color $n$ in $A$, and $d_m$ must have a
neighbor $z_n$ of color $n$, it must be that such a vertex $z_n$
is in $B_1$.  Then $z_n$ sees $x$ since $C$ is a homogeneous set.
%\marginpar{see comment file}
%
%for otherwise $\{z_n, d_m, x, u_m, d_1\}$ induces an $F_1$
%(recall that $u_m$ has no neighbor of color $n$).
%
Likewise $z_n$ sees $y$. But then $\{d_m, z_n, x, y, u_n, u_m,
d_1, d_2\}$ induces an $F_{13}$.  So Claim~(\ref{eq:dmk}) holds.

\begin{equation}\label{eq:dma}
\mbox{Every neighbor of $d_m$ in $A$ is adjacent to all of $N(d_m)\cap
B_1$.}
\end{equation}
For suppose that some neighbor $a$ of $d_m$ in $A$ is not adjacent
to some vertex $y$ in $N(d_m)\cap B_1$.  Vertex $y$ is not in $C$
since $C$ is homogeneous.  So $y$ is adjacent to $u_m$.  Then $a$
sees $u_m$, for otherwise $\{d_1, a, u_m, d_m, y\}$ induces a
$C_5$.  Thus $a\neq u_n$.  Then $u_n$ sees $y$, for otherwise
$\{u_n, d_1, u_m, y, d_m\}$ induces an $F_1$ or $C_5$.  Then $u_n$
misses $d_m$, for otherwise $\{d_1, u_m, u_n, d_m, y, a\}$ induces
a boat ($\overline{C}_6$) or an $F_{10}$.  Since this holds for
every vertex $u_n$ of color $n$ in $A$, and $d_m$ must have a
neighbor $z_n$ of color $n$, it must be that such a vertex $z_n$
is in $B_1$. Since $u_m$ has no neighbor of color $n$, we have
$z_n \in C$. %\marginpar{see comment file}
%
%If $a$ misses $z_n$, then the same argument as with $y$ implies
%that $z_n$ should see $u_n$, which is impossible.
%
Since $C$ is homogeneous, $a$ sees $z_n$.  Note that $a$ sees
$u_n$, for otherwise $\{d_1, a, u_n, d_m, y\}$ induces a $C_5$.
Recall that $z_n$ sees $y$ by (\ref{eq:dmk}) and misses $u_m$ by
the definition of color $n$. But then $\{d_1, a, u_m, u_n, d_m, y,
z_n\}$ induces an $F_{11}$, a contradiction.  So (\ref{eq:dma})
holds.

Vertex $d_m$ must have a neighbor $z_i$ of color $i$ for each
$i\in\{1, \ldots, \ell\}$, and clearly $z_i$ is in $B_1$ since it
is not adjacent to $d_i$.  By (\ref{eq:dmk}) and (\ref{eq:dma}),
every neighbor of $d_m$ (other than $z_i$) is adjacent to $z_i$.
It follows that $z_1, \ldots, z_{\ell}$ are b-vertices, a
contradiction to (\ref{eq:b2b}). This completes the proof of the
main theorem. $\Box$

%%%%%% SECTION 6
%%%%%%
\section{Optimizing b-perfect graphs}

In this section we describe polynomial time algorithms to find an
optimal coloring and a largest clique of a b-perfect graph.

Suppose a graph $G$ is assigned an arbitrary coloring. We want to
find a way to reduce the number of colors used to hopefully obtain
a better coloring of $G$. The notion of b-vertices can be used for
this purpose.  If there is a color $c$ with no b-vertex then we
can eliminate $c$ from our coloring as follows. For each vertex
$x$ of color $c$, give $x$ a color that is missing in the
neighborhood of $x$. We may repeat this process until every color
has a b-vertex, thus obtaining a b-coloring of $G$. We will call
the above algorithm the {\it b-greedy} (coloring) algorithm. It is
easy to see that the b-greedy algorithm can be implemented in
polynomial time. If $G$ is a b-perfect graph, then the b-greedy
algorithm will deliver an optimal coloring since $b(G)=\chi(G)$.

Our notion of b-perfect graph is thus analogous to Chv\'atal's notion
of perfectly orderable graph \cite{chv}.  On a perfectly ordered
graph, the greedy algorithm delivers an optimal coloring.  The
recognition of perfectly orderable graphs is NP-complete \cite{mid};
in comparison, the recognition of b-perfect graphs can be done in
polynomial time, since our main result above is that b-perfect graphs
are characterized by forbidding as induced subgraphs twenty-two
graphs, which have at most eight vertices.

Now, we consider the problem of finding a largest clique of a
b-perfect graph.  First, we need establish some preliminary results.

\begin{lemma}\label{lem:struc1}
Let $G$ be a b-perfect graph that contains a $C_5$.  Then either \\
(i) $G$ has two non-adjacent comparable vertices, or \\
(ii) $X_6 = \emptyset$, and $|X_i| = 1$ for $i = 1,2,3,4,5$ (with the
notation of Lemma \ref{lem:c5}).
\end{lemma}
{\it Proof of Lemma \ref{lem:struc1}.} We know that $G$ has the
structure described in Lemma \ref{lem:c5}, and we use the same
notation.  Write $X''_1 = X_1\setminus X'_1, X''_2 = X_2\setminus
X'_2$.  Let $a_1$ be any vertex in $X'_1$.  Suppose that $X''_1$
contains a vertex $x_1$.  By items 6 and 8 of Lemma~\ref{lem:c5},
$x_1$ dominates $a_1$ and we obtain conclusion (i).  Now let us
assume that $X''_1$ is empty, and similarly $X''_4$ is empty.  Any
two vertices of $X'_1$ are non-adjacent (by item 6 of
Lemma~\ref{lem:c5}) and comparable (by item 8); so if $|X_1| \ge
2$ or $|X_4| \ge 2$ we obtain (i) again.  So let $|X_1| =1 $ and
$|X_4| = 1$.  Suppose that $X_6$ contains a vertex $x_6$.  Since
$|X_4| = 1$, $a_1$ has no neighbor in $X_4$ and so is dominated by
$x_6$ and we obtain (i). Thus, $X_6 = \emptyset$.  Now, if $|X_i|
\geq 2$ ($i=2,3,5$), then $X_i$ contains two non-adjacent
comparable vertices, and we have (i) again.  Thus the lemma holds.
\qed

A {\it special boat} is a boat (with the same notation as in
Section~\ref{sec:boats}) such that all $A_i$'s and all $B_i$'s are
cliques.

\begin{lemma}\label{lem:boat2}
Let $G$ be a $C_5$-free b-perfect graph that contains a boat.  Then
either \\
(i) $G$ has a proper homogeneous set that is not a clique, or \\
(ii) $G$ is a special boat.
\end{lemma}
{\it Proof of Lemma \ref{lem:boat2}.} Consider a set $M$ that induces
a largest boat in $G$.  By Lemma~\ref{lem:boat}, $M$ is a homogeneous
set of $G$.  If $M\neq V(G)$, we obtain conclusion (i).  So let $M =
V(G)$.  Since each $A_i$ and $B_i$ with at least two vertices is a
homogeneous set of $G$, either one of them is not a clique, and we
obtain (i), or all are cliques, and we obtain (ii).  Thus the Lemma
holds.  \qed

\begin{lemma}\label{lem:final}
Let $G$ be a b-perfect graph.  Then one of the following holds:
\begin{enumerate}
 \item
 $G$ has two non-adjacent comparable vertices.
 \item
 $G$ has a proper homogeneous set that is not a clique.
 \item
 $G$ is a $C_5$.
 \item
 $G$ is weakly chordal.
 \item
 $G$ is a special boat.
\end{enumerate}
\end{lemma}
{\it Proof of Lemma~\ref{lem:final}}.  Let $G$ be a b-perfect graph,
and suppose that it does not have two non-adjacent comparable
vertices.  First suppose that $G$ contains a $C_5$, and let us use the
same notation as in Lemma \ref{lem:c5}.  By Lemma~\ref{lem:struc1}, we
have $X_6 = \emptyset$, and $|X_i| = 1$ for $i = 1,2,3,4,5$.  Thus the
set $X=X_1 \cup X_2 \cup X_3 \cup X_4 \cup X_5$ induces a $C_5$.  By
Lemma~\ref{lem:c5}, $X$ is a homogeneous set.  If $X \neq V(G)$, we
obtain conclusion~2.  If $X=V(G)$, we obtain conclusion~3.  Now let
$G$ be $C_5$-free.  If $G$ contains a $\overline{C}_6$ (which is a
boat), then, by Lemma~\ref{lem:boat2}, we obtain conclusion~2 or~5.
If $G$ does not contain a $\overline{C}_6$, then it does not contains
any hole or antihole of length at least five (because $G$ contains no
$F_1=P_5$ and $F_{10}=\overline{P}_6$), and we obtain conclusion~4.
\qed

We are now in position to describe a polynomial-time algorithm to find
a largest clique in a b-perfect graph $G$.  First, if $G$ has two
non-adjacent comparable vertices $x,y$, where $x$ dominates $y$, then
$\omega(G) = \omega(G-y)$; thus, we can remove $y$ from consideration
and recursively find a largest clique in $G-y$.  We apply this
reduction as long as possible.  Second, if $G$ has a proper
homogeneous set $H$ that is not a clique, then we recursively find a
largest clique $K$ of $H$.  Let $G'$ be the graph $G-(H-K)$.  Clearly,
every maximal clique of $G$ either is disjoint from $H$ or contains a
maximal clique of $H$; so $\omega(G) = \omega(G')$.  Note that $G'$
does not have two non-adjacent comparable vertices $x,y$ (for
otherwise, it is easy to see that $x,y$ would have the same property
in $G$, a situation which we have already dealt with); so we do not
need to go back to the first step with $G'$.  Finally, if $G$ does not
have the above two properties, then by Lemma~\ref{lem:final}, $G$ is
either a $C_5$, a special boat, or weakly chordal.  It is easy to
determine $\omega(G)$ when $G$ is a special boat (which is the
complement of a bipartite graph), and there are efficient algorithms
to find a largest clique of a weakly chordal graph \cite{hayspi}.  We
can formalize our algorithm as follows.

% \bt xxxx\=xxxx\=xxxx\=xxxx\=xxxx\kill
% {\bf Algorithm} {\sc CLIQUE(G)}\\
% {\bf Input}: A b-perfect graph $G$\\
% {\bf Output}: A largest clique of $G$\\
% \\
% \> 1. If $G$ has two non-adjacent comparable vertices $x,y$ with
% $x$ \\ \> \>  dominating $y$, then
% return the clique produced by CLIQUE($G-y$)  \\
% \\
% \> 2. If $G$ has a homogeneous set $H$ that is not a clique then \\
% \> \> Let $K$ be the clique returned by CLIQUE($H$) \\
% \> \> Return the clique produced by CLIQUE($G-(H-K)$) \\
% \\
% \> 3. If $G$ is a $C_5$ then return an edge of $G$ \\
% \\
%  \> 4. If $G$ is weakly chordal then \\
%  \> \> Return a largest clique of $G$ produced by the algorithm in
% \cite{hayspi}\\
% \\
% \> /* Now is $G$ is a special boat */ \\
% \> 5. Return (as described below) a largest clique of $G$ \\
% \\
% \et
% %

\noindent
{\bf Algorithm} {\sc CLIQUE(G)}\\
{\bf Input}: A b-perfect graph $G$\\
{\bf Output}: A largest clique of $G$
\begin{enumerate}
    \item % 1
If $G$ has two non-adjacent comparable vertices $x,y$ with $x$
dominating $y$, then return the clique produced by CLIQUE($G-y$);
\item % 2
If $G$ has a proper homogeneous set $H$ that is not a clique then \\
  Let $K$ be the clique returned by CLIQUE($H$) \\
  Return the clique produced by CLIQUE($G-(H-K)$);
\item % 3
If $G$ is a $C_5$ then return a clique of size two of $G$;
\item % 4
If $G$ is weakly chordal then return a largest clique of $G$ produced
by the algorithm in \cite{hayspi};
 \item % 5
 /* Now is $G$ is a special boat */  \\
Return (as described below) a largest clique of $G$.
\end{enumerate}

In Step 5, the special boat is the complement of a bipartite graph.
There are well-known algorithms (see, for example, \cite{chv2}) for
finding a largest clique of the complement of a bipartite graph.
However, we can directly find a largest clique in a special boat given
the sets $A_i, B_i$ (with the same notation as in
Section~\ref{sec:boats}).  Indeed, the largest clique of $G$ is,
clearly, one of the sets $A$, $B$, $A_1\cup B_1$, \ldots, $A_q\cup
B_q$.  So the question is how to find the sets $A_i, B_i$.  When the
algorithm reaches Step 5, we know that $G$ must contain a
$\overline{C}_6$ because $G$ contains no $P_5$, $C_5$,
$\overline{P}_6$ and is not weakly chordal.  By Lemmas~\ref{lem:boat}
and~\ref{lem:boat2}, every boat extends into a special boat containing
all vertices of $G$.  Thus, starting with the $\overline{C}_6$, we can
extend it into a special boat with sets $A_0, A_1, \ldots, A_q, B_0,
B_1, \ldots, B_q$ as desired.  Clearly, this can be done in polynomial
time.

Now, we show that our algorithm can be implemented in polynomial time.
Clearly, Step 1 can be performed in polynomial time.  Considering Step
2, there are many efficient algorithms to find a homogeneous set in a
graph $G$.  The most efficient ones are based on the theory of {\it
modular decomposition}.  This theory is rich and complex, and we
recall here the relevant facts only.  A \emph{module} is defined to be
any homogeneous set $M$ such that every homogeneous set $H$ satisfies
either $H\subseteq M$, $M\subseteq H$, or $M\cap H=\emptyset$.  Note
that $V(G)$ and each singleton $\{v\} \subseteq V(G)$ is a module.
For every module $M$ of size at least two, let $M_1, \ldots, M_h$ be
those modules of $G$ that are properly included in $M$ and are maximal
with that property; then $M_1, \ldots, M_h$ form a partition of $M$;
they are called the \emph{children} of $M$.  The child relation
defines a tree, which is called the \emph{modular decomposition tree}
of $G$.  Note that the root of the tree is the module $V(G)$ and the
leaves of the tree are the singleton modules.  Here is an important
property of every module $M$. (Let $G[M]$ denotes the subgraph of $G$ induced by $M$.) \\
Property (*): If $G[M]$  is not connected, then the children of
$M$ are the vertex-sets of the components of $G[M]$; if
$\overline{G}[M]$ is not connected, then the children of $M$ are
the vertex-sets of the components of $\overline{G}[M]$; if $G[M]$
and $\overline{G}[M]$ are connected, and $H$ is any homogeneous
set of $G$ that is properly
included in $M$, then $H$ is included in a child of $M$.  \\
Now Algorithm {\sc Clique} can be implemented so as to return a
maximum clique $K_M$ of every module $M$ of $G$, starting from the
leaves up to the root, as follows.  If $M$ is a leaf, let the
algorithm return $K_M = M$.  Now suppose that $M$ is not a leaf, that
its children are $M_1, \ldots, M_h$, and that the algorithm has
already produced cliques $K_{M_1}, \ldots, K_{M_h}$.  If $G[M]$ is not
connected, then let $K_M$ be the largest of $K_{M_1}, \ldots,
K_{M_h}$.  If $\overline{G}[M]$ is not connected, then let $K_M$ be
the union of $K_{M_1}, \ldots, K_{M_h}$.  If $G[M]$ and
$\overline{G}[M]$ are connected, then consider the graph $G'_M =
G[K_{M_1}\cup\cdots\cup K_{M_h}]$.  By Property (*), graph $G'_M$ does
not have a homogeneous set that is not a clique.  Moreover, as
observed earlier, $G'_M$ does not have two non-adjacent comparable
vertices.  Thus, by Lemma~\ref{lem:final}, $G'_M$ is either a $C_5$, a
boat or a weakly chordal graph.  So we can compute a maximum clique of
$G'_M$ in polynomial time, and the algorithm returns such a clique as
$K_M$.

There are efficient (but conceptually complex) algorithms that compute
the modular decomposition tree of any graph with $n$ vertices and $m$
edges in time $O(n+m)$, see \cite{mccspi,TCHP}.  Moreover, the modular
decomposition tree has at most $2n$ nodes (including the leaf nodes).
This ensures that Step~2 of the Algorithm is performed at most $n$
times.  Consequently, Steps~3, 4 and~5 are performed at most $n$ times
and Algorithm {\sc Clique} runs in polynomial time.

\end{document}